\documentclass[cernpreprint, atlasdraft=False, texlive=2020, UKenglish, texmf]{atlasdoc}
 
\usepackage{atlaspackage}
\usepackage[backref=false]{atlasbiblatex}
\usepackage{tabularx,booktabs}
\newcolumntype{Y}{>{\centering\arraybackslash}X}
\usepackage[process,jetetmiss]{atlasphysics}
 
\usepackage{atlascontribute}
\AtlasDOI{10.1140/epjc/s10052-022-10785-0}
\AtlasJournalRef{Eur. Phys. J. C 82 (2022) 988}
\graphicspath{{logos/}{figures/}}
 
\usepackage{ANA-EXOT-2020-02-PAPER-defs}
 
\usepackage{multirow, makecell}

\AtlasTitle{Search for type-III seesaw heavy leptons in leptonic final states
in \(pp\) collisions at \( \sqrt{s} = 13\,\text{TeV} \) with the ATLAS detector}
 
\hypersetup{pdftitle={Search for type-III seesaw heavy leptons}, pdfauthor={The ATLAS Collaboration}}
 
\AtlasAbstract{
A search for the pair production of heavy leptons as predicted by
the type-III seesaw mechanism is presented. The search uses proton--proton  
collision data at a centre-of-mass energy of 13\,TeV, corresponding
to \( 139\,\text{fb}^{-1} \) of integrated luminosity recorded by the ATLAS
detector during Run 2 of the Large Hadron Collider.\@
The analysis focuses on final states with three or four electrons
or muons from the possible decays of new heavy leptons via intermediate
electroweak bosons.
No significant deviations above the Standard Model expectation
are observed; upper and lower limits on the heavy lepton
production cross-section and masses are derived respectively.
These results are then combined
for the first time with the ones already published by ATLAS using the channel with two leptons in the final state.
The observed lower limit on the mass of the type-III
seesaw heavy leptons combining two, three and four lepton
channels together is \SI{910}{\text{GeV}}  at the
\SI{95}{\%} confidence level.  
}
 
\author{The ATLAS Collaboration}
 
\AtlasRefCode{EXOT-2020-02}

\AtlasJournal{\EPJC}
 
\PreprintIdNumber{CERN-EP-2021-211}

\hypersetup{pdftitle={ATLAS document},pdfauthor={The ATLAS Collaboration}}
\begin{document}
 
\maketitle
 
\tableofcontents

\section{Introduction}
\label{sec:intro}
 
Neutrino physics presents one of the biggest puzzles yet to be
addressed in modern particle physics. The extremely small values of the neutrino masses compared to
the masses of the other fermions appear unnatural in the Standard Model
(SM)~\cite{Esteban:2018azc}. The seesaw
mechanism~\cite{Minkowski:1977sc,yanagida1979proc,glashow1980proceedings,gell1980supergravity,Mohapatra:1979ia}
provides an elegant way to give a very small mass $m_\nu$ to each of the SM
neutrinos by introducing a heavy Majorana neutrino with mass
$M$. The spontaneously broken electroweak (EW) symmetry explains the neutrino mass
as a Yukawa coupling.
Three types of seesaw mechanisms have been proposed and their
phenomenology can be tested at collider experiments. The \typeIIIseesaw ~\cite{Foot:1988aq}  introduces at
least one extra fermionic $\mathrm{SU(2)_L}$ triplet field coupled to
EW gauge bosons. These heavy charged and neutral leptons
can in principle be produced by EW processes at the Large Hadron Collider (LHC).
 
Type-III seesaw  heavy-lepton searches have already been performed in
various decay channels by both the ATLAS and CMS collaborations. In
\RunOne, ATLAS excluded heavy leptons with masses below
\SI{335}{\GeV}~\cite{EXOT-2014-07}  using final states containing two
light leptons (electrons or muons) and two jets. This mass limit was then
improved  to \SI{470}{\GeV}, still using \RunOne data, by adding the
three-lepton channel~\cite{EXOT-2014-08} as suggested in Ref.~\cite{delAguila:2008cj1}. Using the full \RunTwo data sample of
proton--proton collisions at \(\rts = \SI{13}{\TeV}\),  the CMS
Collaboration has excluded heavy-lepton masses up to
\SI{880}{\GeV}~\cite{CMS-EXO-19-002}  by analysing three- and four-lepton final
states, while ATLAS has excluded heavy-lepton masses up to
\SI{790}{\GeV}~\cite{EXOT-2018-33} by using only the two-lepton-plus-jets final
state. The analysis presented in this paper searches for
a \typeIIIseesaw heavy
lepton in three- and four-lepton final states. For the first time, a combination with the
two-lepton-plus-jets final state is performed, giving a significant improvement in the
sensitivity of the analysis.
 
\begin{figure}[tbp]
\centering
\subfloat[]{
\includegraphics[width=0.4\textwidth]{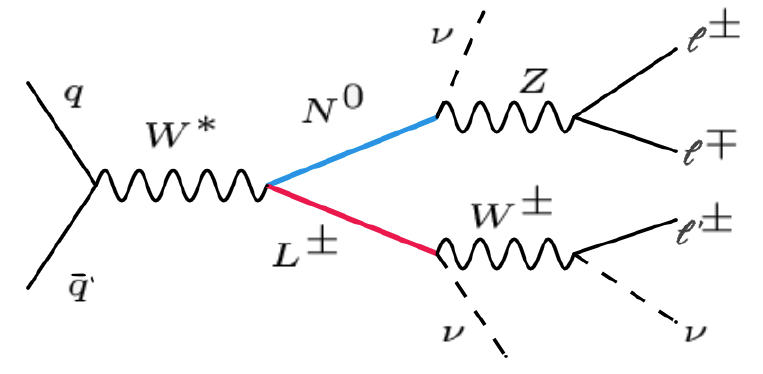}
\label{fig:feynman3l_4la}
}
\subfloat[]{
\includegraphics[width=0.4\textwidth]{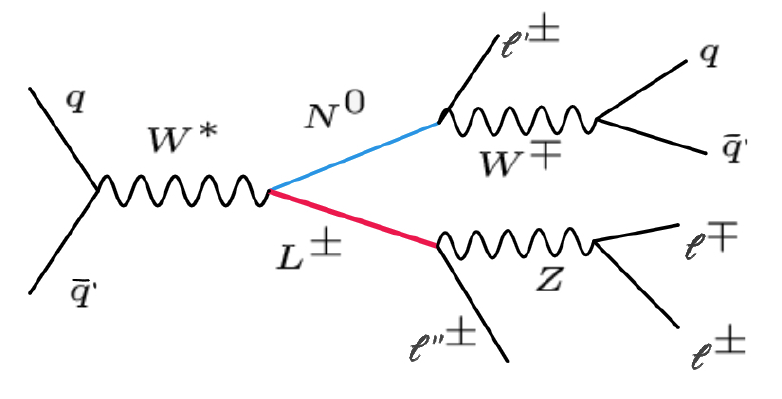}
}\\
\subfloat[]{
\includegraphics[width=0.4\textwidth]{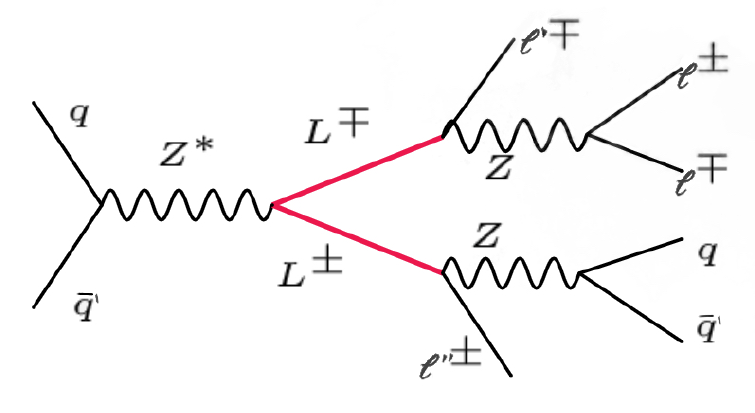}
}
 
\caption{Examples of Feynman diagrams for the considered \typeIIIseesaw model~\cite{Biggio:2011ja}
producing three- and four-lepton final states.
}
\label{fig:feynman3l_4l}
\end{figure}

The  \typeIIIseesaw model targeted in this search is described
in Ref.~\cite{Biggio:2011ja}. It assumes the pair production of the
neutral Majorana ($\Nz$)
and charged ($\Lpm$) heavy leptons proceeds via the
$s$-channel production of virtual EW gauge bosons. \Nz pairs are not
produced because \Nz has $T_3 = Y = 0$ and thus does not couple to the
$\Zboson$~\cite{delAguila:2008cj1,Strumia:2006db}. The production cross-section depends only on the masses of the $\Nz$ and $\Lpm$, which are assumed to be
degenerate as the mass splitting
due to electroweak radiative corrections is expected to be smaller
than \SI{\sim 200}{\MeV} 
~\cite{Arhrib:2009mz}.  The decays allowed in this model are $\Lpm\rightarrow H\ell^\pm,Z\ell^\pm,W^\pm\nu$ and
$\Nz \rightarrow Z\nu, H\nu, W^\pm \ell^\mp$, where the SM leptons can be of any flavour, i.e.\ \( \ell = e,\mu,\tau\).
The branching ratios \(\mathcal{B}_\ell \) for the heavy-lepton decays into $\ell$ plus one SM boson  are
determined by the parameters  \( V_\ell \)  which govern mixing between the new heavy leptons and the SM leptons. Current bounds on \( V_\ell \) can be found in Ref.\cite{Das:2020uer}. In this analysis,
we assume the \textit{democratic} scenario, where the three mixing parameters are equal, so that \(\mathcal{B}_e =
\mathcal{B}_\mu = \mathcal{B}_\tau = 1/3 \). The branching
ratios $\mathcal{B}_{\Zboson}$, $\mathcal{B}_{\Wboson}$, $\mathcal{B}_{\Hboson}$
for heavy-lepton decays into any SM lepton plus $Z$, $W$ or $H$, namely
\(\Lpm,\Nz\rightarrow \Zboson,\Wboson^\pm,\Hboson \)\,,
are independent of the mixing parameters. For \Nz masses
larger than
a few times the \Hboson mass, as considered in this analysis, the decays into different SM bosons
are independent of the heavy-lepton mass,
and therefore \(2 \mathcal{B}_\Hboson \simeq 2 \mathcal{B}_\Zboson \simeq
\mathcal{B}_\Wboson \simeq 1/2 \). Examples of Feynman diagrams
in three- and four-lepton final states are shown in
\Fig{\ref{fig:feynman3l_4l}}.  These events are characterised by the production of two
SM bosons ($VV$, $VH$ or $HH$,  where $V= \Wboson, \Zboson$)  and two charged leptons or neutrinos in the final state. The Majorana nature of \Nz allows
final states with four leptons and non null total lepton electrical
charge as shown in \Fig{\ref{fig:feynman3l_4l}(b)}. This analysis
focuses on events with high light-lepton multiplicity, including light leptons from  \( \tau \)-lepton decays.
 
This paper is structured as follows. The ATLAS detector is described in
\Sect{\ref{sec:detector}}, the data and simulated events used in the analysis are outlined
in \Sect{\ref{sec:data-sim}}, and the event reconstruction procedure is
detailed in \Sect{\ref{sec:definitions}}. The analysis strategy and
background estimation are presented
in \Sects{\ref{sec:regions}}{\ref{sec:background}}, respectively.
The systematic uncertainties are described in \Sect{\ref{sec:systematics}}.
Finally, results and their statistical interpretation are presented in
\Sect{\ref{sec:results}}, followed by the conclusions in \Sect{\ref{sec:conclusion}}.
 
\FloatBarrier

 
\newcommand{\AtlasCoordFootnote}{
ATLAS uses a right-handed coordinate system with its origin at the nominal
interaction point (IP) in the centre of the detector and the \(z\)-axis
along the beam pipe. The \(x\)-axis points from the IP to the centre of the
LHC ring, and the \(y\)-axis points upwards. Polar coordinates
\((r,\phi)\) are used in the transverse plane, \phi being the azimuthal
angle around the \(z\)-axis. The pseudorapidity is defined in terms of the
polar angle \( \theta \) as \( \eta = -\ln \tan(\theta/2) \).
Angular distance is measured in units of
\( \Delta R \equiv \sqrt{{(\Delta\eta)}^{2} + {(\Delta\phi)}^{2}} \).}
 
\section{ATLAS detector}
\label{sec:detector}
 
The ATLAS detector~\cite{PERF-2007-01} at the LHC is a multipurpose particle
detector with a near-\( 4\pi \) coverage in solid angle around the
collision point and a cylindrical
geometry\footnote{\AtlasCoordFootnote} coaxial with the beam
axis. It consists of
an inner tracking detector surrounded by a thin superconducting
solenoid providing a 2 T magnetic field,
electromagnetic and hadronic calorimeters, and a muon spectrometer with superconducting toroidal magnets.

The inner detector (ID) provides charged-particle tracking in the
range \( \abseta < 2.5 \) and, going outwards from the beam pipe,
is composed of a high-granularity silicon pixel detector that
typically provides four measurements per track,
the first hit normally being in the insertable B-layer installed
before Run~2~\cite{ATLAS-TDR-19,PIX-2018-001}, a silicon microstrip
tracker, and a
transition radiation tracker that covers the region up to
\(\abseta = 2.0\).
 
The calorimeter system covers the pseudorapidity range \(|\eta| < 4.9\).
Within the region \(|\eta|< 3.2\), electromagnetic calorimetry is
provided by barrel and
endcap high-granularity lead/liquid-argon (LAr) calorimeters,
with an additional thin LAr presampler covering \(|\eta| < 1.8\)
to correct for energy loss in material upstream of the calorimeters.
Hadron calorimetry is provided by the steel/scintillator-tile calorimeter,
segmented into three barrel structures within \(|\eta| < 1.7\), and
two copper/LAr hadron endcap calorimeters.
The solid angle coverage is completed with forward copper/LAr and
tungsten/LAr calorimeter modules
optimised for electromagnetic and hadronic energy measurements respectively.

The muon spectrometer (MS) instruments the outer part of the
detector and is composed of high-precision tracking
chambers up to  \( \abseta =
2.7 \) and fast detectors for triggering up to \( \abseta =
2.4 \). The MS is  immersed in a magnetic field produced by
three large superconducting air-core toroidal magnets with eight coils
each.
 
A two-level trigger system is used to select  events that are of
interest for the ATLAS physics programme~\cite{TRIG-2016-01}.
The first-level trigger is implemented in hardware and reduces the
event rate to below \SI{100}{\kHz}. A software-based trigger further
reduces this to a recorded event rate of approximately \SI{1}{\kHz}.

An extensive software suite~\cite{ATL-SOFT-PUB-2021-001} is used in the reconstruction and analysis of real and simulated data, in detector operations, and in the trigger and data acquisition systems of the experiment.

\section{Data and simulated events}
\label{sec:data-sim}
 
This analysis uses data  collected in proton--proton collisions at   \(\rts = \SI{13}{\TeV}\)
with proton bunches colliding every \SI{25}{\ns}. After requiring that
all ATLAS subdetectors collected  high-quality data and were operating
normally~\cite{DAPR-2018-01}, the total integrated luminosity amounts to \SI{139}{\ifb}.
The uncertainty in the combined 2015--2018 integrated luminosity is 1.7\%~\cite{ATLAS-CONF-2019-021},
obtained using the LUCID-2 detector~\cite{LUCID2} for the primary luminosity measurements.
Events were collected using dilepton triggers selecting pairs of
electrons~\cite{TRIG-2018-05} or muons~\cite{TRIG-2018-01}.
The transverse momentum (\pt) threshold of the unprescaled dilepton trigger
was raised during the data taking, due to the increasing luminosity of the colliding beams,
but was never higher than \SI{24}{\GeV} for the leading electrons and
\SI{22}{\GeV} for the leading muons.
 
Signal and background events
were modelled using different Monte Carlo (MC) generators as listed in
\Tab{\ref{tab:MC}}. The response of the ATLAS
detector was simulated~\cite{SOFT-2010-01} using the $\GEANT4$
toolkit~\cite{Agostinelli:2002hh} and simulated events were reconstructed with the same
algorithms as those applied to data~\cite{ATL-SOFT-PUB-2021-001}. The
\typeIIIseesaw signal model was implemented in the
\MGNLO~\cite{Alwall:2014hca} generator at leading order (\LO)
using \textsc{FeynRules}~\cite{Alloul:2013bka} and the
\textsc{NNPDF3.0lo}~\cite{Ball:2014uwa} parton distribution function
(PDF) set. All decays  of \Lpm and \Nz into the different leptonic flavours
and subsequent decays of the \Wboson, \Zboson and \Hboson are
considered. Matrix element (ME) events were interfaced to
$\PYTHIA\,8.230$~\cite{Sjostrand:2014zea} for parton showering with the
A14 set of tuned parameters~\cite{ATL-PHYS-PUB-2014-021} and the
\textsc{NNPDF2.3lo} PDF set~\cite{Ball:2012cx}.
The signal cross-section and its uncertainty at next-to-leading-order (NLO) plus
next-to-leading-logarithm (NLL) accuracy were calculated from \( \text{SU}(2)\) triplet production
in an electroweak chargino--neutralino model ~\cite{Fuks:2012qx,Fuks:2013vua}. The calculated
cross-sections are compatible
within uncertainties with the \typeIIIseesaw NLO
implementation~\cite{Ruiz:2015zca,Cai:2017mow}. Production cross
section for a  600, 800 and 1000 \GeV \typeIIIseesaw heavy lepton are
$29.6\pm 3.0$, $7.0\pm 0.8$ and $1.97 \pm 0.25$ \textit{fb} respectively.

Simulated SM background samples include  diboson processes, which are the dominant ones, followed by processes
labelled \textit{rare top quark} that include multi-top-quark
production and
top-quark production in association with EW bosons (\( \ttbar V,
\ttbar H,  t \Wboson\Zboson\)). Other
SM simulated samples are triboson ($VVV$), \ttbar, single top, and
\DY (\( \qqbar \rightarrow \Zboson/\gam^{*} \rightarrow \ellell \,
(\ell=e,\mu,\tau) \)) production processes. They are mainly used for the estimation
of reducible backgrounds as described in~\Sect{\ref{sec:background}}. The MEPS@NLO
prescription~\cite{Hoeche:2012yf} was used in the generation of
\DY processes to match the ME
to the parton shower.
The generators used in the MC sample production and
the cross-section calculations used for MC sample normalisations are listed in
\Tab{\ref{tab:MC}}. The normalisation
of the dominant backgrounds, diboson and rare top-quark processes, are extracted from the final likelihood fit,
as described in \Sect{\ref{sec:results}}.
 
\begin{table}[tbp]
 
\begin{center}
\caption{Configurations used for event generation of signal and
most-relevant background processes. For the cross-section,
the order in the strong coupling constant is shown for the
perturbative calculation. If only one parton distribution
function is shown, the same one is used for both the ME and
parton shower generators; if two are shown, the first is used
for the ME calculation and the second for the parton
shower. Tune refers to the set of tuned underlying-event parameters used by the parton
shower generator.  The masses of the top quark  and SM Higgs
boson were set to 172.5~\GeV\ and 125~\GeV, respectively. The samples with negligible impact are mentioned in the table but not discussed in the text.}
\label{tab:MC}
\vspace{0.25cm}


\scriptsize
\resizebox{\textwidth}{!}{
\begin{tabular}{l l c c c c}
\toprule
Process & Generator &  Cross-    & Parton  &PDF & Tune  \\
&           &  section & shower  &set &    \\
\midrule
Type-III seesaw  & & & & & \\
$\Lp \Lm,\Lpm \Nz$ & \MGNLO\cite{Alwall:2014hca} & NLO+NLL  & $\PYTHIA\,8.230$~\cite{Sjostrand:2014zea}
& \textsc{NNPDF3.0lo}~\cite{Ball:2014uwa} \textsc{NNPDF2.3lo}~\cite{Ball:2012cx}
& A14~\cite{ATL-PHYS-PUB-2014-021} \\
 
\midrule
 
Top quark  & & & & & \\
$\ttbar$ & \textsc{Powheg\,Box}\,v2~\cite{Frixione:2007nw,Nason:2004rx,Frixione:2007vw,Alioli:2010xd} & NNLO & $\PYTHIA\,8.230$ & \textsc{NNPDF3.0nnlo}~\cite{Ball:2014uwa} \textsc{NNPDF3.0nlo}~\cite{Ball:2014uwa} & A14 \\
Single $t$ & \textsc{Powheg\,Box}\,v2 & NNLO & $\PYTHIA\,8.230$ & \textsc{NNPDF3.0nnlo} \textsc{NNPDF3.0nlo} & A14 \\
 
\midrule
Rare top quark  & & & &  \\
3$t$, 4$t$ & \MGNLO & LO &  $\PYTHIA\,8.230$ & \textsc{NNPDF3.0lo} & A14 \\
$\ttbar$ + $W/Z/H$, $tWZ$ & \MGNLO & NNLO & $\PYTHIA\,8.230$ &
\textsc{NNPDF3.0nlo} & A14 \\
\midrule
Diboson & & & & & \\
$ZZ$, $WZ$ & $\SHERPA\,2.2.1$~\cite{Bothmann:2019yzt} \& 2.2.2 & NLO & \SHERPA & \textsc{NNPDF3.0nnlo} & \SHERPA default \\
\midrule
Triboson & & & & & \\
$WWW, WWZ, WZZ, ZZZ$ & $\SHERPA\,2.2.1$ \& 2.2.2 & NNLO
& \SHERPA & \textsc{NNPDF3.0nnlo} & \SHERPA default \\
\midrule
 
Drell--Yan & & & & & \\
$\Zboson/\gam^{*}\rightarrow \ee/\mumu/\tautau$ &
$\SHERPA\,2.2.1$ & NLO & \SHERPA & \textsc{NNPDF3.0nnlo}  & \SHERPA default \\

\bottomrule
\end{tabular}
}


\end{center}
\end{table}

Diboson \(VV\) and triboson  \(VVV\) events were
simulated with the
$\SHERPA\,2.2.1$--2 generator~\cite{Bothmann:2019yzt}. Off-shell effects and Higgs boson contributions
were included. ME calculations were matched
and merged with the \SHERPA parton shower based on the Catani--Seymour  
dipole factorisation~\cite{Gleisberg:2008fv,Schumann:2007mg} using the
MEPS@NLO
prescription~\cite{Hoeche:2011fd,Hoeche:2012yf,Catani:2001cc,Hoeche:2009rj}. The
\OPENLOOPS library~\cite{Cascioli:2011va,Denner:2016kdg-fixed}
provided QCD corrections. Diboson samples
were produced with different lepton multiplicities
using MEs at NLO accuracy in QCD for up to one additional parton
and at LO accuracy for up to three additional parton emissions.
Loop-induced \( gg \to VV \)  processes were
generated with LO MEs for emission of up to one additional parton
for both the fully leptonic and semileptonic final states.
Electroweak production of a diboson pair in association with two jets
(\(VVjj\)) was also simulated at LO\@. The PDFs used for the nominal samples
were CT14~\cite{Dulat:2015mca}
and MMHT2014~\cite{Harland-Lang:2014zoa}.
 
Samples of events from \( \ttbar + V\) (where \( V \) stands for \( \gamma^\ast, \Wboson,\Zboson\, \mathrm{and}\, \Hboson \)) and \(t \Wboson \Zboson\) processes were produced using
\MGNLO~\cite{Alwall:2014hca} at NLO accuracy and were
interfaced to the  $\PYTHIA\,8.230$ parton
shower with \textsc{NNPDF3.0nlo}~\cite{Ball:2014uwa} PDFs. The \hdamp parameter, which controls
the matching between the ME and the parton shower, was set
to  \SI{1.5}{\mtop}~\cite{ATL-PHYS-PUB-2016-020},
using a top quark mass of \( \mtop = \SI{172.5}{\GeV}
\).

All simulated events were overlaid with a simulation of multiple \pp interactions
occurring in the same or neighbouring bunch crossings. These  \(pp\) inelastic scattering events were
generated by $\PYTHIA\,8.186$
using the \textsc{NNPDF2.3lo} set of PDFs and the A3 set of tuned
parameters~\cite{ATL-PHYS-PUB-2016-017}. Their effects
are referred to as \textit{pile-up}. The simulated events were reweighted such
that the distribution of the average number of interactions per
bunch crossing is compatible with that observed in the data.
 

\section{Event reconstruction}
\label{sec:definitions}
 
Events considered in this analysis are required to have at least one collision vertex
reconstructed with at least two tracks with  transverse momentum
greater than \SI{500}{\MeV}.  The primary  vertex of the
hard-scattering event is the one with  the largest sum of the
associated tracks' squared transverse momenta. Events have to satisfy
the quality criteria listed in
Ref.~\cite{DAPR-2018-01}, including rejection of events with a large
amount of calorimeter noise or non-collision background.

Electrons are reconstructed by matching a charged-particle track in the ID with an
energy deposit in the electromagnetic calorimeter. Electron candidates
are required to satisfy a \textit{Loose} likelihood-based identification selection~\cite{EGAM-2018-01} and to be in the fiducial
volume of the inner detector, \( \abseta < 2.47 \). The transition region
between the barrel and endcap calorimeters (\( 1.37 < \abseta < 1.52 \))
is excluded because it is partially non-instrumented due to services infrastructure.
The transverse impact parameter \( \dzero \) of the track associated with
the candidate electron must have a significance satisfying \( |\dzero|/\sigma(\dzero) < 5\).
This is required in order to reduce the number of electrons originating from secondary decays.
Similarly, the track's longitudinal impact parameter $z_0$ relative to the primary vertex must
satisfy $|z_0\sin(\theta)|<\SI{0.5}{\mm}$, where $\theta$ is the track's polar angle.
The electron's transverse energy \ET must exceed \SI{10}{\GeV}. After this preselection, to refine the electron quality,
a  \textit{Tight}  likelihood-based
identification selection and a set of \textit{Loose} isolation criteria based on both
calorimetric and tracking information are applied to primarily select electrons
coming from the decays of the heavy leptons or the EW bosons.
 
Track segments in the MS are matched with ID tracks to reconstruct
muons if they
are within the $\eta$ coverage of the ID\@.  Muon
candidates with \pt lower than \SI{300}{\GeV} are required to satisfy the
\textit{Medium}
muon identification requirements, while for high-\pt muons, a specific identification
working point is applied~\cite{MUON-2018-03}. Muon candidates are
required to have $\abseta < 2.5$, a transverse
impact parameter significance of \( |\dzero|/\sigma(\dzero) < 3 \) and
a longitudinal impact parameter value of
\( |z_{0}\sin(\theta)| < \SI{0.5}{mm} \).  The minimum muon \pt is \SI{10}{\GeV}.
After this preselection, an isolation requirement based only on
tracking information is applied.

In this analysis, particle-flow objects~\cite{PERF-2015-09} are formed from energy-deposit clusters in the calorimeters
and tracks measured in the ID but not matched to identified leptons.
Particle-flow objects  are then clustered into
jets using the \antikt algorithm~\cite{Cacciari:2008gp}  with a
radius parameter $R = 0.4$. The measured jet \pt is
corrected for detector effects to measure the particle energy before
interactions with the detector material~\cite{PERF-2016-04}. Energy within jets that is
due to \pileup is estimated and removed by subtracting an amount equal to the mean
pile-up energy deposition density multiplied by the \( \eta
\)--\( \phi \) area of the jet. Pile-up can also produce additional
jets that are identified and rejected by the  jet-vertex tagger
(JVT) algorithm~\cite{PERF-2014-03}, which distinguishes them from jets originating from
the hard-scattering primary vertex. Only jets with transverse energy greater
than \SI{20}{\GeV}  and $|\eta| < 2.4$ are considered.
Jets originating from heavy-flavour quarks are identified with the
MV2c10 multivariate $b$-tagging algorithm using the
\SI{77}{\%} efficiency working point~\cite{PERF-2012-04,ATL-PHYS-PUB-2016-012,ATL-PHYS-PUB-2017-013},
with measured rejection factors of approximately 134, 6 and 22 for
light-quark and gluon jets, \(c\)-jets, and hadronically decaying
\( \tau \)-leptons, respectively.
 
The missing transverse momentum $\vec{p}_{\mathrm{T}}^{\mathrm{miss}}$ (with magnitude \MET)
is calculated as the negative vectorial sum of the \pt of reconstructed jets and leptons in the event.
A `soft term' taking into account tracks associated with the primary vertex but not with any hard object
is then added to guarantee the best performance in a high \pileup environment~\cite{PERF-2016-07}.
The \MET significance \metsig, calculated with a maximum-likelihood ratio method, is used in
\Sect{\ref{sec:regions}} to define the various analysis regions, taking into account the direction of
the   $\vec{p}_{\mathrm{T}}^{\mathrm{miss}}$  and calibrated objects as well as their respective resolutions.

Different objects reconstructed close together in the $\eta$--$\phi$ plane could in principle have originated from
the same primary object. Possible overlaps are resolved by an algorithm that appropriately removes one of the two closely spaced objects to avoid double-counting. If a muon candidate is found to share an ID track with an electron candidate,
the electron candidate is rejected. If two electron candidates
share an ID track, the one with the lower
\pt is rejected. Jets are rejected if they are within $\DeltaR = 0.2$ of a lepton candidate, except if the candidate is
a muon and three or more collinear tracks are found. Finally, lepton candidates that are within $ \DeltaR= 0.4$
of any remaining jet are removed.
 

\section{Analysis strategy}
\label{sec:regions}

Once events have been classified according to the presence of exactly
either three or
four light leptons,\footnote{Leptons are ordered going from the
highest to the lowest momentum, $\ell_1$ to $\ell_3$ ($\ell_4$ for
four-lepton events).}
the two lepton-multiplicity categories are refined with dedicated
selections. Events with larger lepton multiplicities can be
categorized in lower lepton multiplicities regions if one or more leptons escape
detection.
Signal regions (SRs) are defined so as to maximise the significance of
the signal event
count predicted by the targeted model relative to the expected number of SM
background events. SM backgrounds are normalised
by performing a simultaneous fit in the SRs and in dedicated
control regions (CRs). The CRs are defined so as to be
enriched in relevant background processes and depleted in events from
signal processes.
The fit uses a kinematic variable chosen to
optimise the sensitivity to the small cross-sections expected for the signal processes.
Validation regions (VRs), also
depleted in signal events, are used to validate the extrapolation of the SM
background expectations
obtained from the background-only fit to
independent regions kinematically
close to the SRs.  All CRs and VRs are characterised by a signal contamination below 2\%.

Backgrounds are assigned to two broad categories: reducible and irreducible backgrounds.
Reducible backgrounds include leptons from misreconstructed objects such as jets, or from
light- or heavy-quark decays or, in the electron case, photon conversions.
These are called \textit{fake or non-prompt} (FNP) leptons, and events containing at least one such lepton are referred to as the FNP background.
Its contribution is calculated using a data-driven method.
Irreducible backgrounds are produced  by SM processes with three or
four prompt leptons in the final state.
Prompt leptons are leptons produced in the decays of \Wboson bosons, \Zboson bosons, and $\tau$-leptons,
as well as direct decays of the heavy leptons considered as signal in this analysis.
The most important sources are diboson and rare top-quark processes, with
the latter being primarily \ttbar pairs produced in association with an EW or Higgs boson.
For these processes, kinematic distributions are obtained from MC simulation and their normalisation
is extracted from the fit.
Because low-mass heavy leptons have been excluded by previous searches,
this search focuses on higher masses, where signal events are characterised by objects having high momenta.

Details about the three- and four-lepton analysis regions are provided below, while details of the two-lepton analysis regions are
given in Ref.~\cite{EXOT-2018-33}. The analysis regions are all orthogonal to one another.
 
\newcommand{\MTfootnote}{
The generic transverse mass of one or multiple objects $N_{\mathrm{obj}}$ is defined as:
$m_{\mathrm{T}}^2(N_{\mathrm{obj}}) = \left(\sum_i^{N_{\mathrm{obj}}}E_{\mathrm{T},i} + E_{\mathrm{T}}^{\mathrm{miss}}\right)^2- |\sum_i^{N_{\mathrm{obj}}}{\vec{p}_{\mathrm{T},i}} + \vec{p}_{\mathrm{T}}^{\mathrm{miss}} |^2 $.
}

\subsection{Three-lepton channel}
 
\Tab{\ref{tab:3l_regions}} summarises the selection criteria used to define the three-lepton SRs, CRs and VRs.
The  ZL SR is characterised by a
leptonically decaying \Zboson boson, and thus an opposite-sign, same-flavour (OSSF) lepton pair compatible with the \Zboson boson mass
is required. The SM boson from the decay of the other heavy lepton produced in the event, decays hadronically.
Signal events are expected to have a large three-lepton invariant mass ($m_{\ell\ell\ell}$),
and the transverse masses of the two highest-\pt leptons,
$m_\mathrm{T}(\ell_{1})$ and $m_\mathrm{T}(\ell_{2})$, are also expected to be large.\footnote{\MTfootnote}
An additional
requirement is placed on the
angular distance between the leading and subleading leptons to
further increase the signal-to-background ratio in the SRs.

A complementary ZLveto SR, targeting signals involving leptonic decays
of \Wboson bosons and hadronic decays of \Zboson bosons (including those from \Hboson bosons),
is defined by vetoing events containing OSSF lepton pairs compatible with a leptonic decay of an on-shell \Zboson boson
requiring the  invariant mass of the pair to be larger than 115\, \GeV. The \HT variable is defined as the scalar sum of the \pt of all selected objects in the event. In cases where the scalar sum of the \pt is restricted to only a subset of the objects, they are specified.
Signal events are characterised by large \HT
and \met values and by a large value of the scalar sum of the momenta of the
same-sign leptons, denoted by $\HT (\mathrm{SS})$.
Since the presence of SS leptons in this region is mainly due
to rare top events and FNP leptons, the $\HT$ of this
pair is used as discriminating variable looking for values
$\HT(\mathrm{SS}) \ge 300$\GeV. To account for possible hadronic
decays of electroweak bosons from diboson background
sources, an upper limit is placed on the two leading jets invariant
mass $m_{jj}$.
 
Finally, the JNLow SR targets events where the electroweak bosons
decay leptonically, and therefore events with low jet multiplicity,
as in \Fig{\ref{fig:feynman3l_4la}}, are selected. A lower bound is
imposed on the invariant mass of
the OSSF lepton pair ($m_{\ell\ell}(\mathrm{OSSF})$)\footnote{If two
OSSF lepton pairs are present, the requirement is applied to both pairs.} and a large value of the scalar
sum of the \pt of the three leptons, $\HT(\ell\ell\ell)$, is required.
Fake-lepton background is further reduced by requiring
$m_{\mathrm{T}}(\ell_{1})$ and $m_{\mathrm{T}}(\ell_{2})$ to exceed a minimum value.
The angular separation $\Delta R(\ell_1,\ell_2)$ between the two leptons is required to exceed a
minimum value to reduce the FNP contribution.
Overall selection efficiencies for the production of a 800 \GeV \typeIIIseesaw
heavy lepton are 0.29 \%, 0.57\%, 0.41\% for the ZL, ZLveto and JNLow
SR respectively.

SM backgrounds in the three-lepton SRs consist of diboson events,
which contribute
${\sim}60\%$, ${\sim}80\%$ and ${\sim}40\%$ in the ZL,
JNLow and ZLveto regions, respectively. Another background in the ZL and
ZLveto regions originates from rare top-quark processes involving one or more top quarks, which contribute
${\sim}40\%$ and ${\sim}50\%$ of the background in those regions, respectively.
Therefore, a CR targeting the normalisation of the diboson
background is defined by requiring at least two jets and a low transverse
mass for the subleading lepton such that $m_\mathrm{T}(\ell_2)\le \SI{200}{\GeV}$.
 
Two VRs are defined in order to validate background estimates for events containing a \Zboson boson decaying into leptons, both obtained by inverting the
$\DeltaR(\ell_1,\ell_2)$ selection of the ZL SR, and applying additional requirements. The DB-VR also requires a $b$-tag
veto, while in the RT-VR the presence of at least one $b$-tagged jet is required. These VRs validate the predictions and normalisation of diboson and rare top-quark processes respectively. An additional JNLow-VR is obtained from the JNLow
SR by inverting the transverse mass requirement on the leading lepton,
$m_\mathrm{T}(\ell_1)\le \SI{240}{\GeV}$. Moreover, a Fake-VR is
defined by inverting the \metsig selection common to all the other
regions without applying any additional requirement except for lepton
\pt ones. This region is enriched in contributions from FNP
backgrounds and is therefore used to validate them.
 
In the three-lepton channel SRs, the kinematic variable used as the final discriminant to the data is the transverse mass of the three-lepton system.

\begin{table}[htpb]
\begin{center}
\caption{ Summary of the selection criteria used to define
relevant regions in the three-lepton analysis. No selection is
applied when a dash is present in the corresponding cell.
}
\vspace{0.25cm}

\renewcommand{\arraystretch}{1.3}
\setlength\tabcolsep{3.2pt}
\footnotesize
 
\begin{tabularx}{\textwidth}{c | *{8}{Y|}}
\toprule
 
\multicolumn{2}{c |}{ }
& \multicolumn{4}{c |}{\textbf{ZL} }
& \multicolumn{1}{c |}{\textbf{ZLveto} }
& \multicolumn{2}{c |}{\textbf{JNLow} }\\
 
\hline
& Fake-VR & CR & DB-VR & RT-VR &  SR  &  SR &  VR & SR\\
\hline
 
& \multicolumn{8}{  c |}{ $\pt (\ell_1) > 40 \ \GeV$ } \\
 
& \multicolumn{8}{ c |}{ $\pt (\ell_2) > 40 \ \GeV$  } \\
& \multicolumn{8}{ c |}{ $\pt (\ell_3) > 15 \ \GeV$  } \\
\hline
$\metsig$ & $<5$ & \multicolumn{7}{ c |}{ $ \ge 5 $  }  \\

\hline
$N(\mathrm{jet})$
& -
& \multicolumn{5}{c |}{$\geq 2$}
& \multicolumn{2}{c |}{$\leq 1$} \\
\hline
 
$N(\bjet)$
 
& - & - & \multicolumn{1}{c |}{$0$}
& \multicolumn{1}{c |}{$\geq 1$} & - & - & - & - \\
\hline
 
$m_{\ell\ell} \, (\mathrm{OSSF}) ~ [\GeV]$ & -
& \multicolumn{4}{c |}{80--100 }
& \multicolumn{1}{c |}{$\geq 115$}
& \multicolumn{2}{c |}{ $\geq 80$ } \\
\hline

$H_{\mathrm{T}}+\met ~ [\GeV]$
& - & - & - & - & -
& \multicolumn{1}{c |}{$\geq600$}
& - & - \\
\hline
 
$m_{\ell\ell\ell} ~ [\GeV]$
& - & - & \multicolumn{3}{c |}{$\geq 300$}
& \multicolumn{1}{c |}{$\geq 300$}
& - & - \\
\hline
 
$H_{\mathrm{T}}(\mathrm{SS}) ~ [\GeV]$
& - & - & - & - & -
& \multicolumn{1}{c |}{$\geq 300$}
& - & -  \\
\hline
 
$m_{jj} ~ [\GeV]$
& - &-  &-  &-  &-
& \multicolumn{1}{c |}{$< 300$}
& - & - \\
\hline
 
$H_{\mathrm{T}}(\ell\ell\ell) ~ [\GeV]$
& - & - & - & - & - & -
& \multicolumn{2}{c |}{$\geq 230$}  \\
\hline

$m_{\mathrm{T}}(\ell_1) ~ [\GeV]$ &
- & - &   \multicolumn{3}{c |}{$\geq 200$}
&  -
& \multicolumn{1}{c |}{$< 240$} & \multicolumn{1}{c |}{$\geq 240$}  \\
\hline
 
$m_{\mathrm{T}}(\ell_2) ~ [\GeV]$ & -
&   \multicolumn{1}{c |}{$< 200$} &   \multicolumn{3}{c |}{$\geq 200$}
&  -
& \multicolumn{2}{c |}{$\geq 150$}  \\
\hline
 
$\Delta R (\ell_1, \ell_2)$ &
- & -  &  \multicolumn{2}{c |}{$< 1.2 $}   &  \multicolumn{1}{c |}{1.2--3.5}
&  -
&   \multicolumn{2}{c |}{$\geq 1.3$}  \\

\bottomrule
\end{tabularx}


\label{tab:3l_regions}
\end{center}
\end{table}

\subsection{Four-lepton channel}
 
In the four-lepton channel, the momentum requirement on the three leading leptons is the same as in the three-lepton channel; the momentum of the fourth lepton is required to be larger than $\SI{10}{\GeV}$. Events are classified using the sum  of the charges of the
four leptons in the final state: $\sum q_\ell$. The conditions $\sum q_\ell= 0$  and
$|\sum q_\ell|= 2$ identify the \textit{zero charge} (Q0) and
\textit{double charge} (Q2) regions, respectively.
A summary of the selection criteria defining the four-lepton regions
is shown in  \Tab{\ref{tab:4l_regions}}.
Signal events are characterised by  large \HT and
large invariant mass $m_{\ell\ell\ell\ell}$ of the four-lepton system. The presence of possible neutrinos in the final state is taken into account using the  $\HT + \met$ variable;  therefore
$m_{\ell\ell\ell\ell}\ge\SI{300}{\GeV}$ and $\HT + \met \ge \SI{300}{\GeV}$
are required in both the Q0 and Q2 signal regions.
Apart from the common lepton \pT requirements, these are the only kinematic selections applied in the Q2 SR,
which has less background than the Q0 SR since it is
very rare for a SM process to produce a doubly charged final state.
To reduce the \textit{ZZ$^\ast$} contribution in
the Q0 SR, no more than one OSSF
lepton pair in an event is allowed to be
compatible with a leptonic \Zboson decay defined by the invariant mass window  80--100\,\GeV.
Background in the Q0 SR is further reduced by requiring $\metsig \ge
5$. Overall selection efficiencies for the production of a 800 \GeV \typeIIIseesaw
heavy lepton are 0.14 \% and  0.11\% for the Q0 and Q2 SR respectively.
 
Two CRs and two VRs are defined in the zero-charge Q0 kinematic space. A DB-CR targeting diboson
backgrounds is built by requiring a $b$-jet veto and defining an invariant mass window for
the four-lepton system ($\SI{170}{\GeV}\le m_{\ell\ell\ell\ell} <
\SI{300}{\GeV}$). A RT-CR targeting rare top-quark background
is obtained by requiring at least two $b$-tagged jets and
$m_{\ell\ell\ell\ell} <  \SI{500}{\GeV}$. To ensure orthogonality to the CRs,
the VRs require events to have exactly one $b$-jet. To increase the
contributions of diboson and rare top-quark backgrounds in the VRs, $m_{\ell\ell\ell\ell}$
must satisfy  $\SI{170}{\GeV}\le m_{\ell\ell\ell \ell} <
\SI{300}{\GeV}$ in DB-VR and $ \SI{300}{\GeV} \le m_{\ell\ell\ell\ell} <
\SI{500}{\GeV}$ in  RT-VR\@. The RT-VR also requires $\HT + \met \ge \SI{400}{\GeV}$ and $\metsig \ge 5$.
 
The main sources of background in the Q2 signal region are diboson or rare top-quark events where the electric charge of one of the
electrons is mismeasured. As mentioned above, the only additional kinematic selections used to
define the Q2 SR are that both $\HT + \met$ and the four-lepton invariant mass must exceed $\SI{300}{\GeV}$. A dedicated Q2 VR
is obtained by requiring $m_{\ell\ell\ell\ell} < \SI{200}{\GeV}$ or $\HT + \met < \SI{300}{\GeV}$ in order to validate both the
diboson and FNP background estimates.

In the four-lepton channel SRs, the kinematic variable $\HT+\met$ is used as the final discriminant to fit to the data.

\begin{table}[htpb]
\begin{center}
 
\caption{ Summary of the selection criteria used to define relevant regions in the four-lepton analysis. $N_\Zboson$ is the number of leptonically reconstructed \Zboson bosons,
using opposite-sign same-flavour leptons. No selection is
applied when a dash is in the corresponding cell.
}
\label{tab:4l_regions}
\vspace{0.25cm}

\renewcommand{\arraystretch}{1.3}
\setlength\tabcolsep{3.2pt}
\footnotesize

\begin{tabularx}{\textwidth}{c | *{8}{Y|}}
\toprule
 
& \multicolumn{5}{c |}{\textbf{Q0} } & \multicolumn{2}{c |}{\textbf{Q2} } \\
 
\hline
 
& DB-CR & RT-CR & DB-VR & RT-VR & SR & VR & SR \\
\hline
& \multicolumn{7}{  c |}{ $\pt (\ell_{1,2}) > 40 \ \GeV$~~\, } \\
 
& \multicolumn{7}{ c |}{ $\pt (\ell_3) > 15 \ \GeV$  } \\
& \multicolumn{7}{ c |}{ $\pt (\ell_4) > 10 \ \GeV$  } \\
\hline
$| \sum{q_\ell} |$ & \multicolumn{5}{c |}{ 0 } & \multicolumn{2}{c |}{ 2 }    \\
\hline
$N(\bjet)$ & 0 & $\geq 2$ & 1 & 1 & 0  & - & -  \\
\hline

\multirow{2}{*}{$m_{\ell\ell\ell\ell} ~ [\GeV]$} & \multirow{2}{*}{170--300} & \multirow{2}{*}{$<500$} & \multirow{2}{*}{170--300}   & \multirow{2}{*}{300--500} & \multirow{2}{*}{$\geq 300 $}   & $<200$ & \multirow{2}{*}{$\geq 300$} \\
& \multicolumn{1}{c|}{} & \multicolumn{1}{c|}{} & \multicolumn{1}{c|}{} & \multicolumn{1}{c|}{} &  \multicolumn{1}{c|}{} & OR &    \\
\cline{1-6}\cline{8-8}
$H_{\mathrm{T}}+\met ~ [\GeV]$ & - & - & - & $\geq 400 $ & $\ge 300 $  & $<300$ &  $\ge 300 $  \\
\hline
 
$N_{Z}$ & - & - & - & - & $\leq 1$ & - & -  \\
\hline
 
$\metsig$ &  - & - & - & $\geq 5 $  & $\geq 5$ & - & -  \\
 
\bottomrule
\end{tabularx}


\end{center}
\end{table}


 
\section{Background composition and estimation}
\label{sec:background}
 
The background estimation techniques used in the analysis, combining simulations and data-driven methods common to all channels, are discussed in this section.
 
Irreducible-background predictions are obtained directly from simulations, but normalisation of diboson and rare top-quark
processes is obtained from the fit.
To avoid double-counting between background estimates derived from MC simulation and
the data-driven reducible-background
predictions, a specific check is performed: events from irreducible-background MC samples
are considered only if generator-level prompt leptons can be matched to their reconstructed counterparts.

There are two sources of reducible background: events in which at least one lepton charge is misidentified and FNP leptons. The former source is
relevant only in the Q2 four-lepton signal region where an event
in the Q0 category can migrate to the Q2 category if the charge of one of the leptons is mismeasured.
 
Charge misidentification for
muons is well described by the simulation and occurs only for
high-momentum muons where detector misalignments degrade the muon
momentum resolution. However, electrons are more susceptible to charge misidentification, as a combination of effects from bremsstrahlung and photon conversions that might not
be adequately described by the detector simulation.
Correction factors (scale factors) accounting for charge misreconstruction are
applied to the simulated background events. They
are derived by comparing the charge misidentification probability
measured in data with the one in simulations and are parameterised as
function of \pt and \eta. The charge
misidentification probability is extracted
by performing a likelihood fit in a dedicated \Zee data sample, as
described in Ref.~\cite{EGAM-2018-01}. The charge misidentification
probability increases from ${\sim}10^{-4}$ to
${\sim}10^{-1}$ with increasing \pt and $|\eta|$.

FNP leptons are produced by secondary decays
of light- or heavy-flavour mesons into light leptons embedded within
jets. Although the \(b\)-jet veto and lepton isolation significantly
reduce the number of FNP leptons, a fraction
still satisfy the selection requirements.
Significant components of FNP electrons arise from photon conversions and from jets
that are wrongly reconstructed as electrons.
MC samples are not used to estimate these background sources because
the simulation
of jet production and hadronisation has large intrinsic uncertainties.
 
Instead, the FNP background
is estimated with a data-driven approach, known as the \textit{fake factor} (FF)
method, as described in Ref.~\cite{EXOT-2012-25}. 
The FF is measured in dedicated
FNP-enriched regions where the events satisfy single-lepton triggers
without isolation requirements and have low \met,
no \(b\)-jets, and only one reconstructed lepton that
satisfies the lepton identification preselection described in
\Sect{\ref{sec:regions}}. In these regions, two kinds of leptons are identified, $\mathcal{L}$ leptons satisfy looser object selection criteria than the ones used to identify the leptons considered in the analysis regions, which are here named $\mathcal{T}$ leptons.
Electron and muon FFs are
then defined as the ratio of the number of $\mathcal{T}$ leptons to the number
of $\mathcal{L}$ leptons, and are parameterised as functions of \pt and
\eta. The FNP background
is then estimated in the SRs by applying the FF as
an event weight
in a template region defined with the same selection criteria as the
corresponding SRs, except that at least one of the leptons must be an $\mathcal{L}$ lepton but not a $\mathcal{T}$ one.
 
The prompt-lepton contribution is subtracted from the template region by
using the irreducible-background MC samples to estimate the prompt-lepton
contamination in the adjacent regions~\cite{EGAM-2018-01}.
The Fake-VR, as defined in \Tab{\ref{tab:3l_regions}}, is used to validate
the data-driven FNP-lepton estimate.


\section{Systematic uncertainties}
\label{sec:systematics}
 
Uncertainties affect several aspects of this analysis. Experimental
uncertainties related to the trigger selection, lepton reconstruction,
identification, momentum measurement and isolation selection affect
both the global selection efficiency and the shape of the kinematic
distributions used in the fit. The main contributions come from the electron
selection efficiency and the sagitta resolution of the muon spectrometer.
Uncertainties are estimated mainly from  $\Zboson\rightarrow \ell \ell$
and $\jpsi \rightarrow \ell \ell$ processes~\cite{EGAM-2018-01,MUON-2018-03}.
Uncertainties in the jet energy scale and
resolution are evaluated from MC simulations and data, using
multi-jet, \Zjets
and $\gamma$\,+\,jets events~\cite{ATLAS:2020cli}; they are estimated to be
less than \SI{2}\% in the range of jet transverse momentum of interest and
affect both the selection efficiency measurement and the kinematic
distributions used in the fit.
Uncertainties in the
\btag efficiencies are evaluated from data using dileptonic \ttbar events
\cite{FTAG-2018-01}. They
are estimated to range from \SI{8}\% at low momentum to \SI{1} \% at high momentum.
These uncertainties affect the analysis region selection
efficiencies. The \MET measurement uncertainties are estimated from data-to-MC
comparison in \Zmm events without jets
as described in \Refn{\cite{ATLAS-CONF-2018-023}}. The \MET uncertainties affect
both the selection efficiencies and the kinematic distributions used in the
fit. The charge misidentification uncertainty
affects only the Q2 analysis regions. The uncertainty in the
charge-misidentification scale factor is estimated to be less
than 10\% from a comparison between same-sign dielectron data and MC events,
with the same electron selection as used in this analysis, and
with $|m_{ee}-m_\Zboson|<\SI{10}{\GeV}$ as described in
\Refn{\cite{EGAM-2018-01}}. The uncertainty in the \pileup simulation,
derived from a comparison of data with simulation, is also taken
into account~\cite{ATLAS-CONF-2014-018}. The limited size of the MC
samples is taken into account as an additional uncertainty.
 
The FNP background uncertainty comes from the modelling and normalisation
of the prompt leptons subtracted in the  FF estimation. The origin of the
FNP background is then varied by selecting slightly modified FNP-enriched regions,
where the FF is measured. Variations of the FNP-enriched regions
are, for example, obtained by varying the jet multiplicity
requirement and the \MET
selection. 
The resulting uncertainty of the FF depends on the lepton momentum and
pseudorapidity and ranges from 5\% to 40\% for electrons and from
10\% to 30\%  for muons.

Theoretical uncertainties affect both the signal and background
predictions. For both, the uncertainties from missing higher orders
are evaluated by independently varying the
QCD factorisation and renormalisation scales in the matrix element
by up to a factor of two~\cite{Bothmann:2016nao}.
The PDF uncertainties are evaluated using the LHAPDF
toolkit~\cite{Buckley:2014ana} and the PDF4LHC prescription~\cite{Butterworth:2015oua}.
An additional
uncertainty of \SI{10}\% is added to the diboson cross-section to take into
account variations in the level of data-to-MC agreement for \VV
processes in different jet multiplicity regions.
For rare top-quark backgrounds, uncertainties in the
$\ttbar \Wboson $ cross-section are evaluated to be $\pm 50\%$, while
for  the $\ttbar \Hboson$ cross-section the uncertainty from varying the
QCD factorisation and renormalisation scales is $^{+5.8}_{-9.2}\%$, with another $\pm 3.6\%$
from PDF+\alphas variations.
Since the yields  of the rare top-quark and diboson backgrounds are derived from
the likelihood fit to the data in the CRs, the systematic variations
have little impact on the final yields of the background predictions
in the CR and SR.


\section{Statistical analysis and results}
\label{sec:results}
 
The \textsc{HistFitter}~\cite{Baak:2014wma} statistical package is
used to fit the predictions to the data in the CRs and SRs. The fit considers the $m_{\mathrm{T},3\ell}$ and $\HT+\met$ distributions for the three- and four-lepton channels, respectively, calculating lower limits on the heavy-lepton mass with a test statistic based on the binned profile likelihood in the asymptotic approximation~\cite{Cowan:2010js}, whose validity is tested using a pseudo-experiment approach. The lower limits on the mass are calculated at 95\% confidence level (CL) and the binning is chosen to
optimise the sensitivity to  signal. The various
components of the background
predictions are validated in the corresponding VRs.
Background and signal contributions are modelled by a product of independent Poisson
probability density functions representing the
likelihood of the fit. Systematic uncertainties are modelled by Gaussian probability density functions centred on
the pre-fit prediction of the nuisance parameters, with widths
that correspond to the magnitudes of these uncertainties.
Four different fitting procedures are
performed: the three-lepton channel on its own, the four-lepton
channel on its own, the three- and four-lepton channels combined, and finally the two-, three- and four-lepton channels
combined, where results for the two-lepton channel are taken from
Ref.~\cite{EXOT-2018-33}. All the contributions from the experimental
uncertainties in the lepton, jet and
\MET selections and reconstruction,  pile-up simulation, background
simulation,
theoretical calculations and
irreducible-background estimates are considered correlated among the
different multiplicity channels in multi-channel fits.

After a background-only likelihood fit in the CRs, the three- and four-lepton channel
diboson normalisation factors are found to be $0.80\pm 0.09$ and
$1.08\pm0.03$, respectively. The normalisation and shape
of the $m_{\mathrm{T},3\ell}$ and $\HT+\met$ distributions are
validated in the ZL DB-VR and Q0 DB-VR, respectively, by comparing data and SM
expectations after the fit.
The rare top-quark contribution normalisation is estimated to be
$1.3\pm0.2$ in the four-lepton channel
Q0 RT-CR and is then extrapolated to all the SRs. The
background modelling is validated in the ZL RT-VR and
Q0 RT-VR for the three- and four-lepton channels,
respectively.

\begin{figure}[htpb]
\centering
\includegraphics[width=\textwidth]{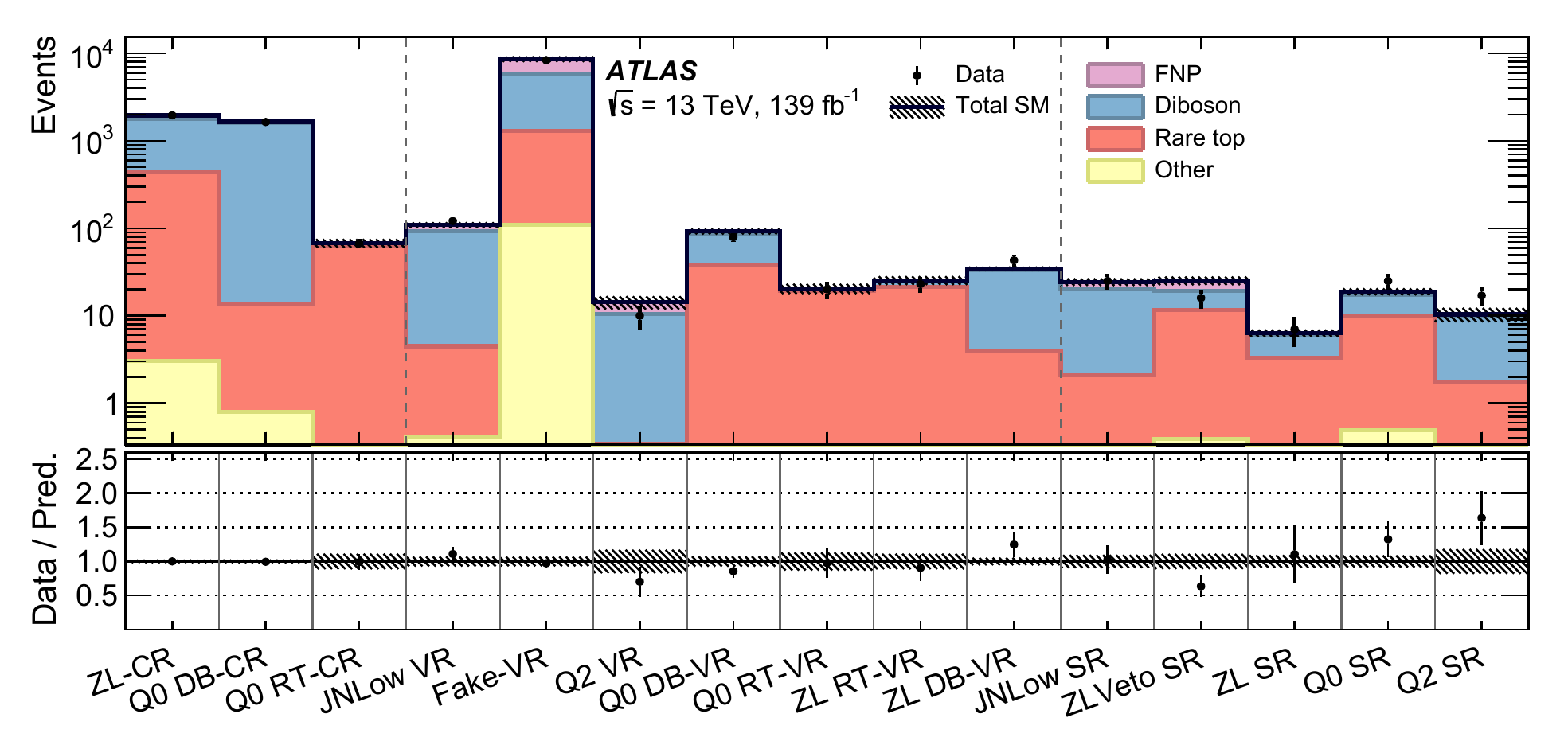}
\caption{Observed and expected event yields in the CRs, VRs and SRs for the three- and four-lepton channels after the fit procedure described in the text. \textit{Diboson}
indicates background from diboson processes. \textit{Rare
top} indicates  background from \( \ttbar + V \) and  \(t \Wboson \Zboson\)
processes. \textit{FNP} includes the background from fake or non-prompt leptons.
\textit{Other} indicates all the other considered backgrounds that
contribute less than 2\%. The hatched bands
include systematic uncertainties with the correlations between
various background sources taken into account.
The lower panel shows the ratio of the observed data to the predicted
SM background after the likelihood fit. }
\label{fig:34Regions}
\end{figure}

Event yields after the likelihood fit for the analysis regions in the three- and four-lepton channels
are shown in \Fig{\ref{fig:34Regions}}. 
Good agreement within
statistical and systematic uncertainties
between data and SM predictions is observed in all regions,
demonstrating the validity of the background
estimation procedure as shown in \Tab{\ref{tab:TriFourLeptonSR}}.


\begin{table}
\begin{center}
\caption{Observed data and background yields in the three- and four-lepton signal regions after the background-only fit in the combined three- and four-lepton regions; the combination of statistical and systematic uncertainties have been reported.
}
\vspace{0.25cm}{\footnotesize
\setlength{\tabcolsep}{3.5pt}

\begin{tabular}{lS[table-format=2.3]
@{\,}@{$\pm$}@{\,} S[table-format=1.3] S[table-format=2.2]
@{\,}@{$\pm$}@{\,} S[table-format=1.2] S[table-format=2.2]
@{\,}@{$\pm$}@{\,} S[table-format=1.2] S[table-format=2.2]
@{\,}@{$\pm$}@{\,} S[table-format=1.2] S[table-format=2.4]
@{\,}@{$\pm$}@{\,} S[table-format=1.4] S[table-format=1.4]}
 
\toprule
& \multicolumn{6}{c}{Three-lepton signal regions} &   \multicolumn{4}{c}{Four-lepton signal regions} \\
\midrule
& \multicolumn{2}{c}{ZL SR}            & \multicolumn{2}{c}{ZLveto SR}            & \multicolumn{2}{c}{JNLow SR}   & \multicolumn{2}{c}{Q0 SR}            & \multicolumn{2}{c}{Q2 SR}                   \\
\midrule
Data          & \multicolumn{2}{l}{~~7}              & \multicolumn{2}{l}{16}              & \multicolumn{2}{l}{25}          & \multicolumn{2}{l}{25}              & \multicolumn{2}{l}{17}                           \\
\midrule
 
Total background         & 6.25 & 0.52  & 25.2 & 2.8  & 24.4 & 2.3    & 19.0 & 1.6  & 10.3 & 1.9     \\
\midrule
Diboson         & 2.62 & 0.27  & 7.64 & 0.95  & 18.0 & 2.1   & 7.70 & 0.78  & 8.5 & 1.6     \\
Rare top         & 3.2 & 0.5  & 11.2 & 1.7  & 1.82 & 0.32  & 9.4 & 1.4  & 1.63 & 0.35     \\
Fakes         & 0.29 & 0.05  & 5.98 & 0.85  & 4.3 & 0.5   & 1.37 & 0.36  & 0.07 & 0.37     \\
Other          & 0.113 & 0.015  & 0.36 & 0.12  & 0.33 & 0.03    & 0.49 & 0.04  & 0.1001 & 0.0098  \\
 
\bottomrule
\end{tabular}
}
\label{tab:TriFourLeptonSR}
\end{center}
\end{table}

\begin{figure}[tbp]
\centering
\subfloat[]{
\includegraphics[width=0.45\textwidth]{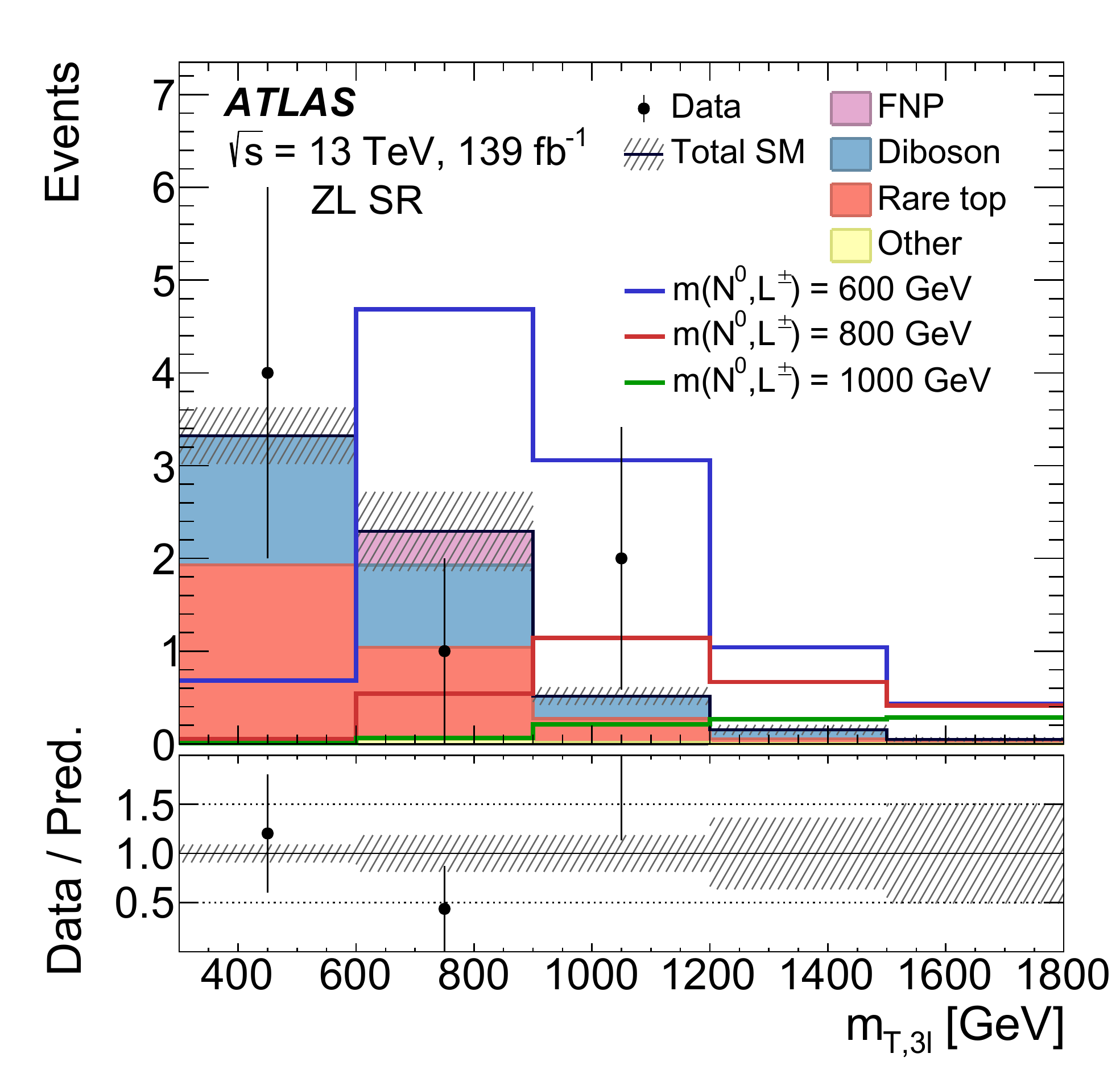}
\label{fig:Post_Signal_3L_ZL}
}
\subfloat[]{
\includegraphics[width=0.45\textwidth]{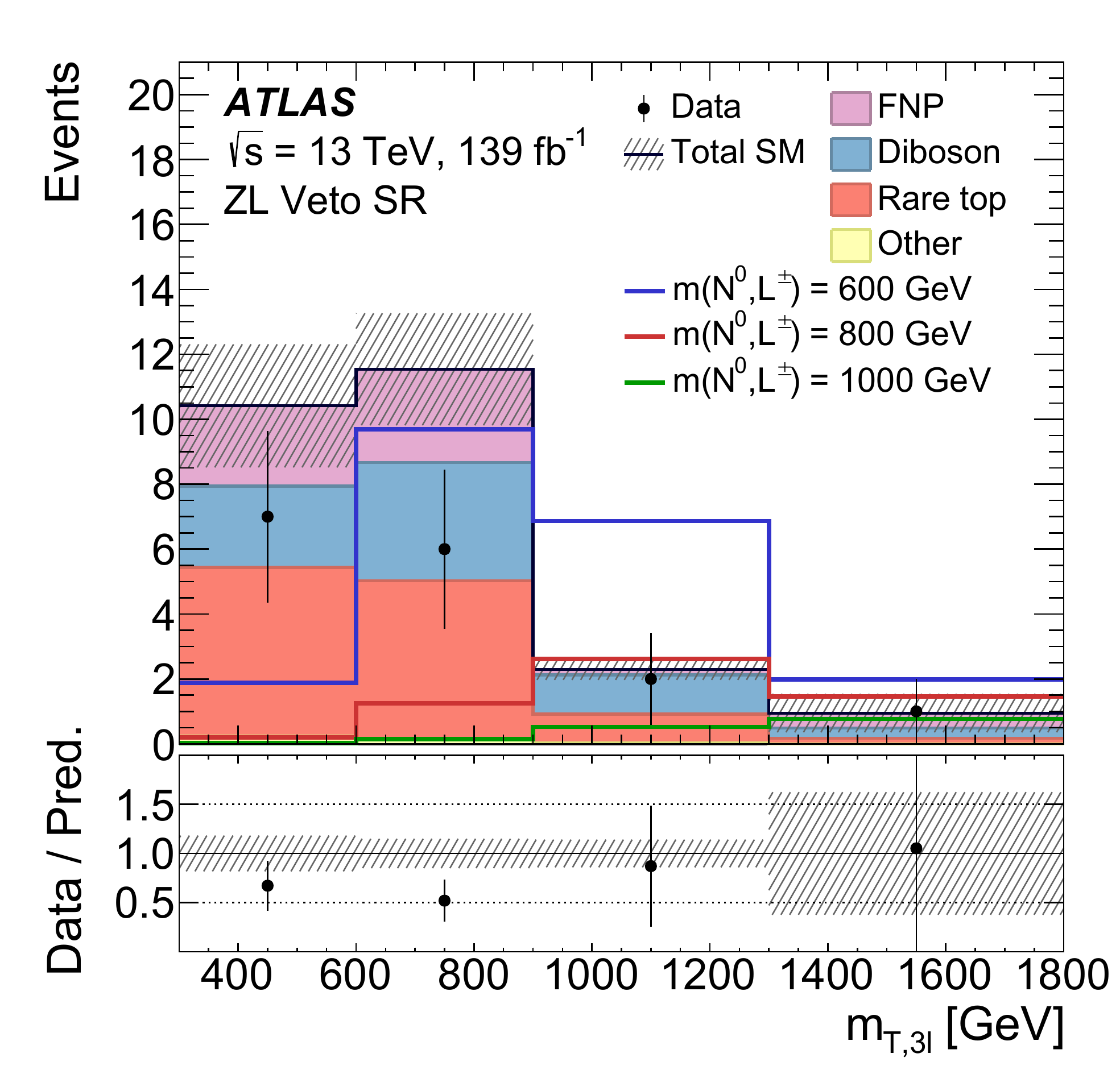}
\label{fig:Post_Signal_3L_ZLVeto}
}\\
\subfloat[]{
\includegraphics[width=0.45\textwidth]{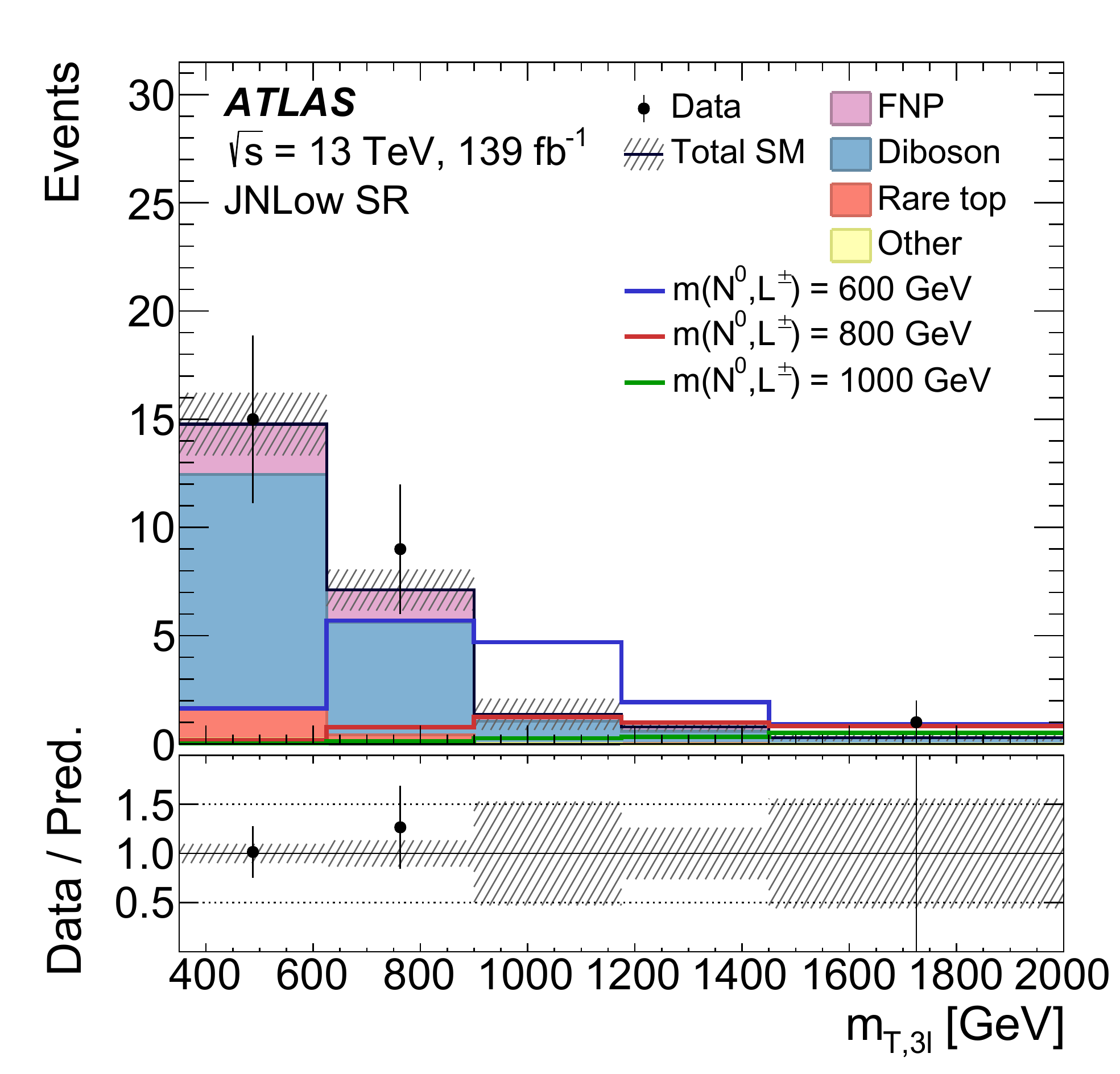}
\label{fig:Post_Signal_3L_JNLow}
}

\caption{Distributions of $m_{\mathrm{T},3\ell}$ in the three-lepton signal
regions after the combined fit:
\protect\subref{fig:Post_Signal_3L_ZL} the ZL signal region,
\protect\subref{fig:Post_Signal_3L_ZLVeto} the ZLveto signal
region and
\protect\subref{fig:Post_Signal_3L_JNLow} the JNLow signal
region. The coloured lines
correspond to signal samples with the \Nz and \Lpm mass values stated in
the legend. The hatched bands include all statistical and systematic
post-fit uncertainties with the correlations between various background
sources taken into  account. The lower panel shows the
ratio of the observed data to the predicted SM background. The
last bin in the distributions contains the overflows.
}
\label{fig:Post_Signal_3L}
\end{figure}

\begin{figure}[tbp]
\centering
\subfloat[]{
\includegraphics[width=0.45\textwidth]{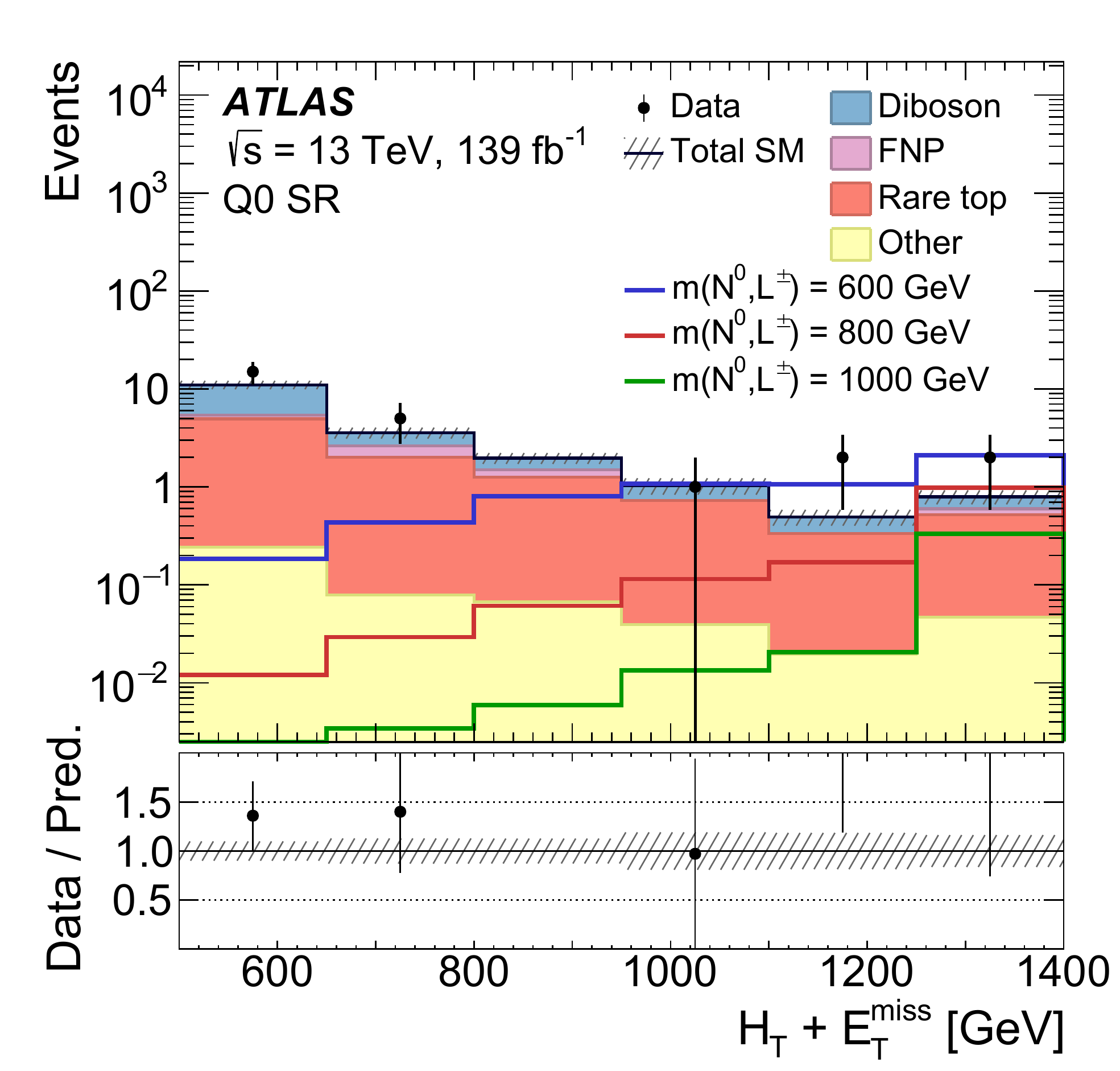}
\label{fig:Post_Signal_4L_HH}
}
\subfloat[]{
\includegraphics[width=0.45\textwidth]{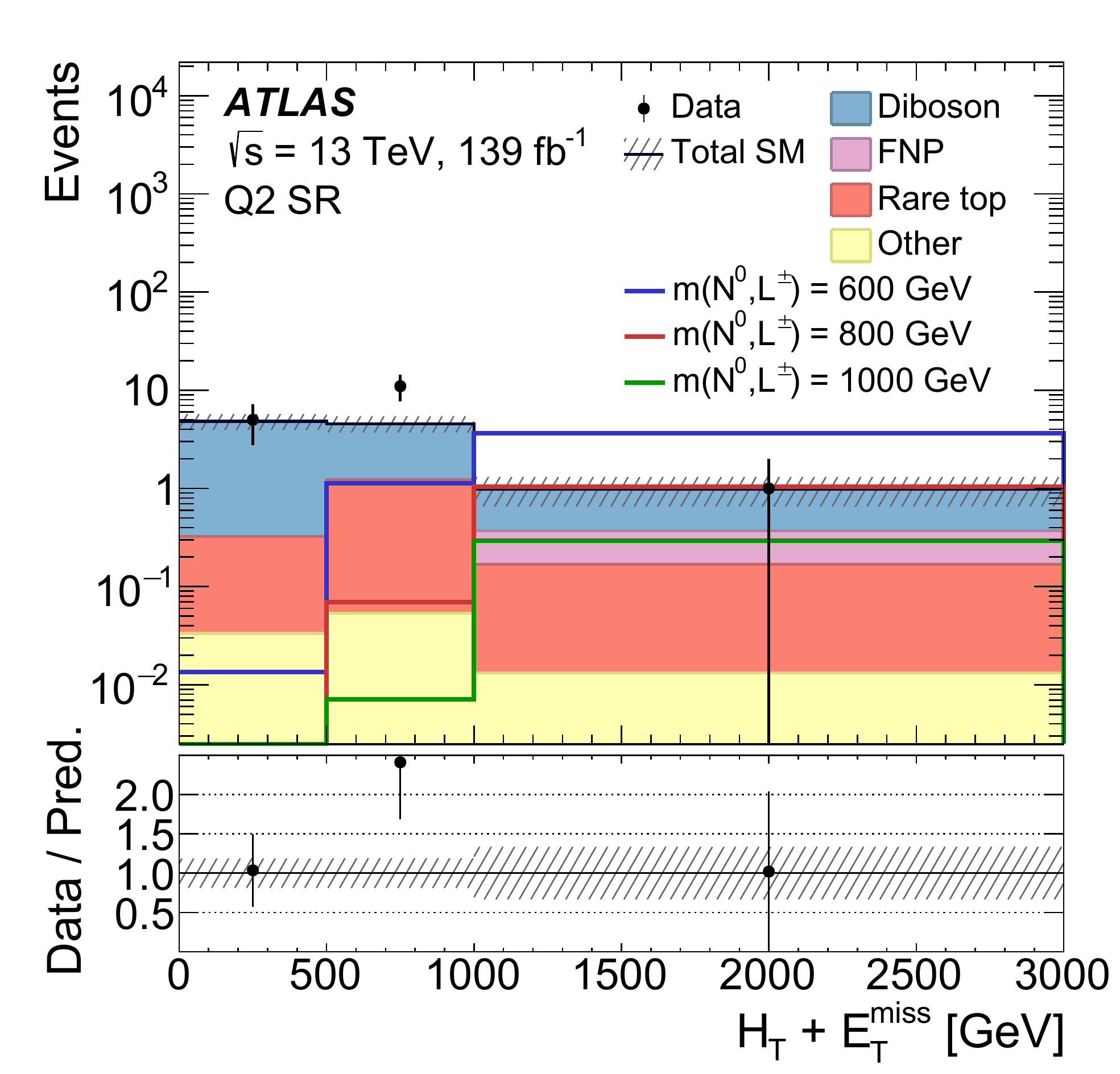}
\label{fig:Post_Signal_4L_OD}
}

\caption{Distributions of $\HT + \met$ in the four-lepton signal
regions after the combined fit:
\protect\subref{fig:Post_Signal_4L_HH} the Q0 signal region where
the sum of lepton charges is zero and
\protect\subref{fig:Post_Signal_4L_OD} the Q2 signal region where
the sum of lepton charges is $\pm 2$. The coloured lines
correspond to signal samples with the \Nz and \Lpm mass values stated in
the legend. The  hatched bands include all statistical and systematic
post-fit uncertainties with the correlations between various background
sources taken into  account.
The lower panel shows the
ratio of the observed data to the predicted SM background. The
last bin in the distributions contains the overflows.
}
\label{fig:Post_Signal_4L}
\end{figure}

\begin{figure}[tbp]
\centering
\subfloat[]{
\includegraphics[width=0.40\textwidth]{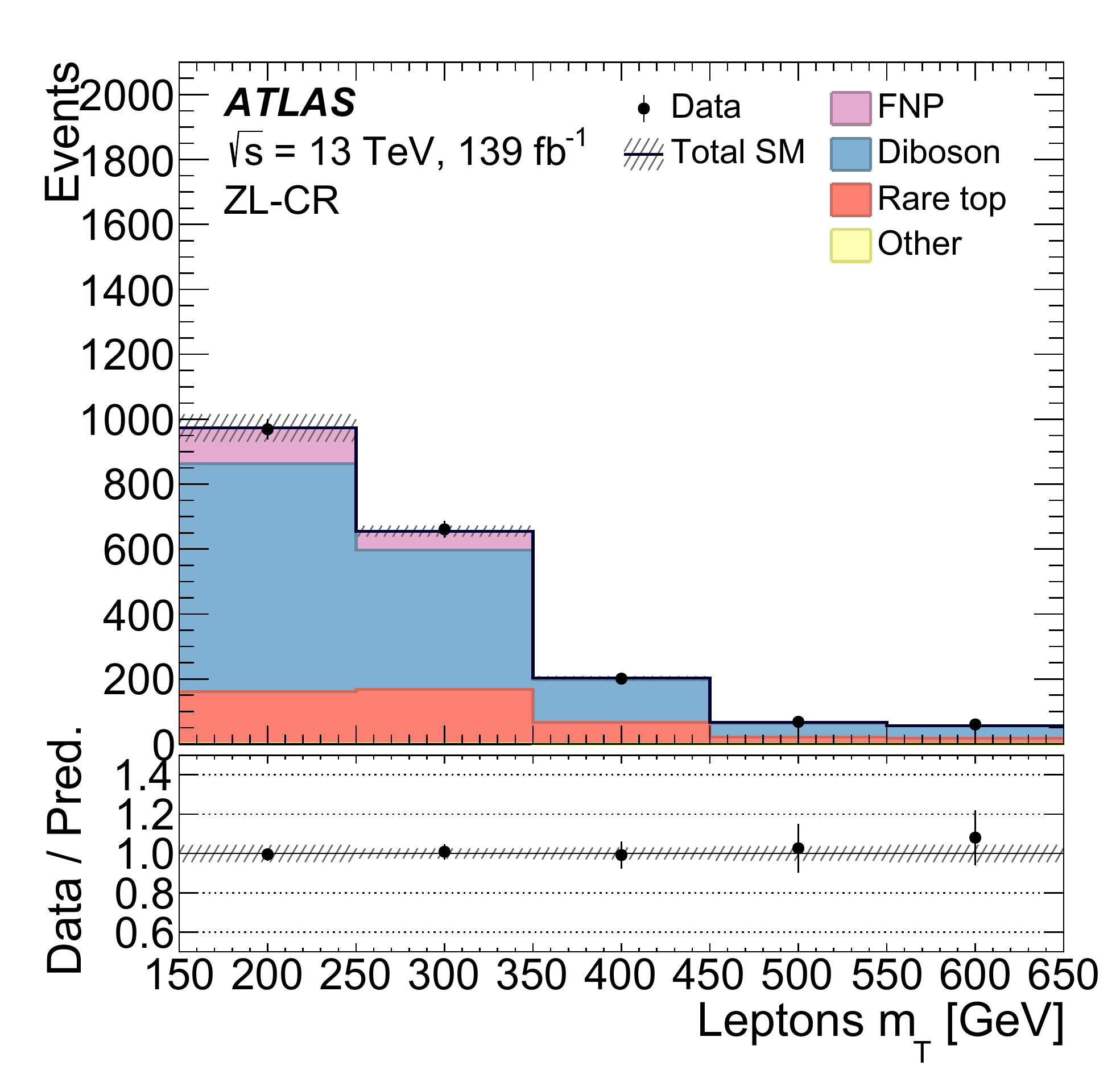}
\label{fig:zlcr}
}
\subfloat[]{
\includegraphics[width=0.40\textwidth]{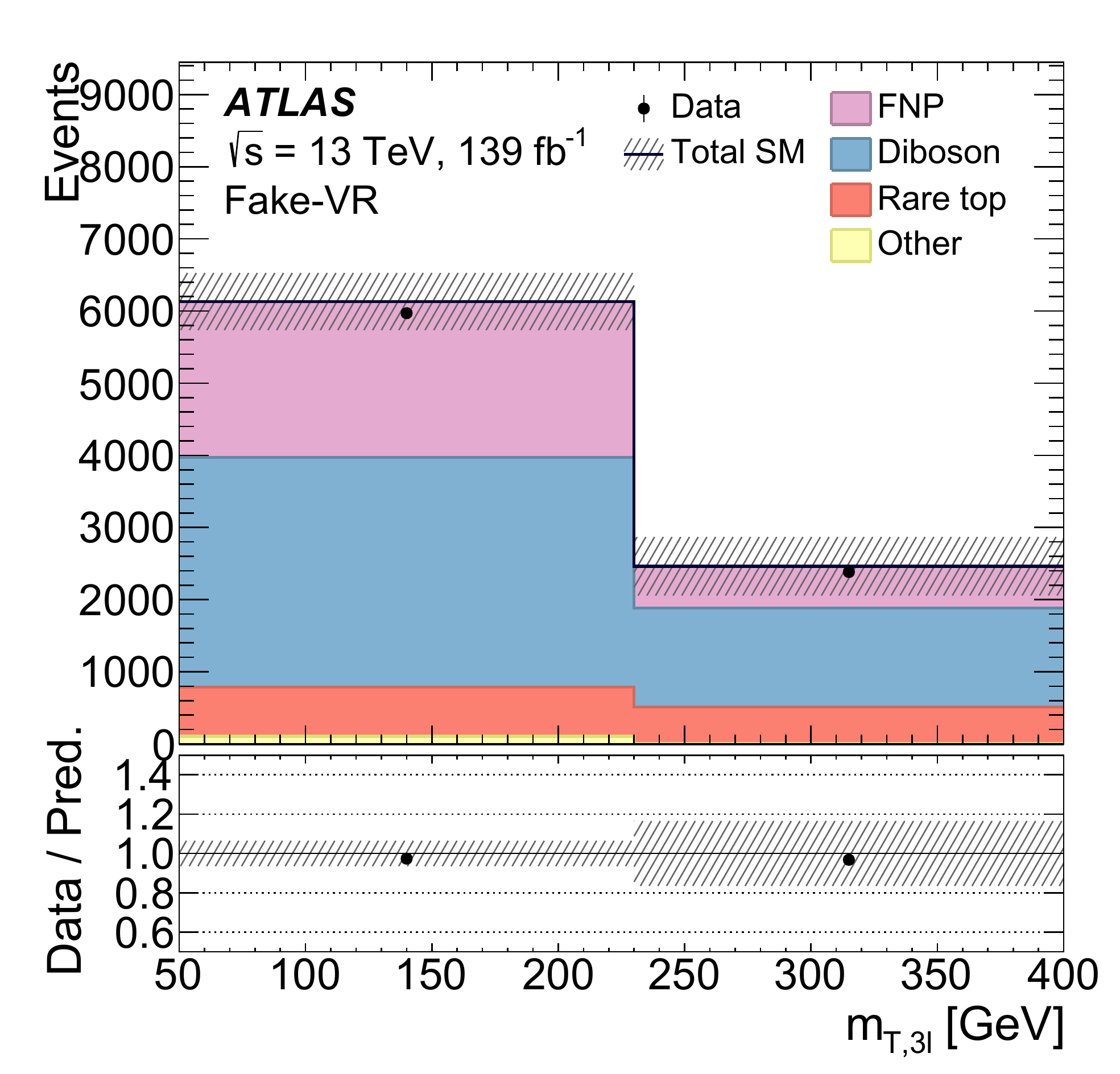}
\label{fig:fvr}
}
\\
\subfloat[]{
\includegraphics[width=0.40\textwidth]{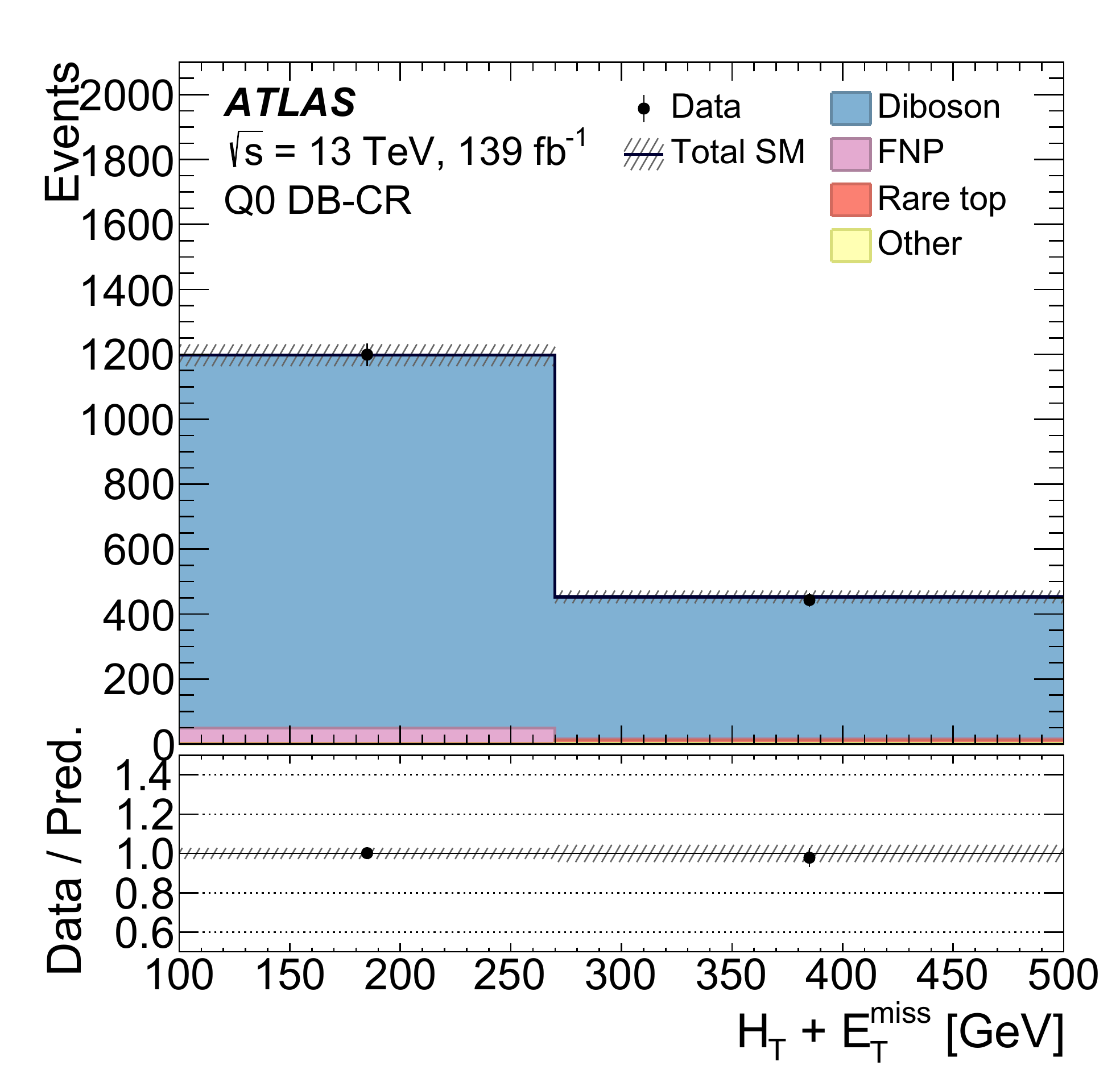}
\label{fig:q0dbcr}
}
\subfloat[]{
\includegraphics[width=0.40\textwidth]{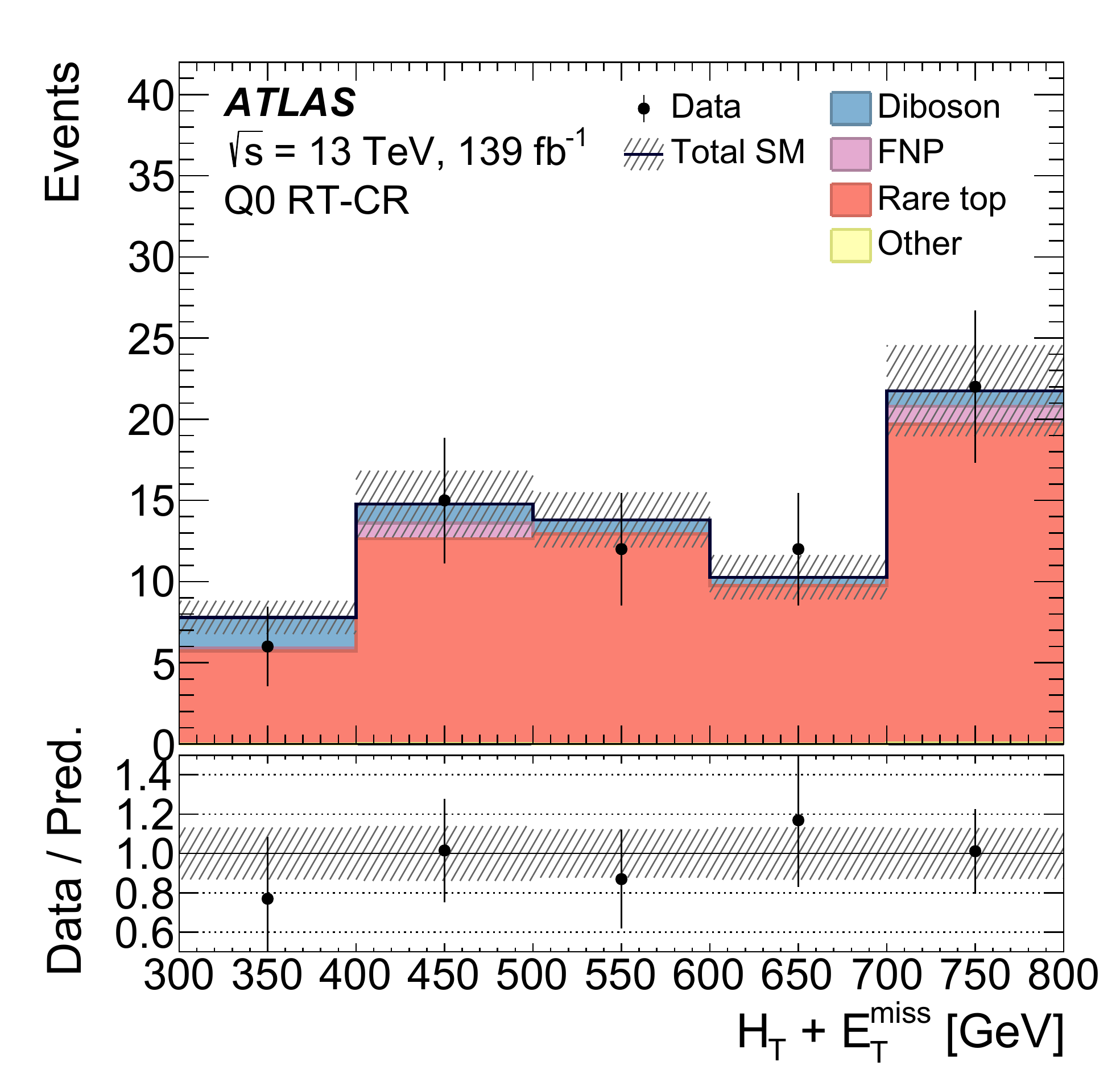}
\label{fig:rtcr}
}

\caption{Distributions of $m_{\mathrm{T},3\ell}$ in the three-lepton control and validation regions
\protect\subref{fig:zlcr} ZL-CR and \protect\subref{fig:fvr} fake-VR,
and of $\HT+\met$ in the four-lepton control regions \protect\subref{fig:q0dbcr} Q0 DB-CR
and \protect\subref{fig:rtcr} Q0 RT-CR  after the combined fit.
The simulated signal contribution was found to
be below 2\% and is not shown in the figure.  The hatched
bands include all statistical and systematic
post-fit uncertainties with the correlations between various background
sources taken into  account. The lower panel shows the
ratio of the observed data to the predicted SM background. The
last bin in the distributions contains the overflow.
}
\label{fig:crs_lep}
\end{figure}

The post-fit distributions of the $m_{\mathrm{T},3\ell}$ and $\HT+\met$ variables used in the likelihood
fit in the three- and four-lepton channels are shown in
\Figs{\ref{fig:Post_Signal_3L}}{\ref{fig:Post_Signal_4L}}, respectively, for the
signal regions, with the binning used in the fit.
After the fit,
the compatibility of the data and the expected background is assessed.
Good agreement is observed. The \textit{p}-values\footnote{The \(p\)-value is defined as
the probability of observing an excess at least as large as the one observed in data, in the absence of signal.},
evaluated using the distributions in \Figs{\ref{fig:Post_Signal_3L}}{\ref{fig:Post_Signal_4L}},
are 0.38, 0.090  and 0.25 for
the three-, four- and the combined three-and-four lepton channels, respectively. \Fig{\ref{fig:crs_lep}} shows the distributions of these discriminating variables in some of the control and validation regions.


\begin{figure}[htpb]
\centering
\includegraphics[width=\textwidth]{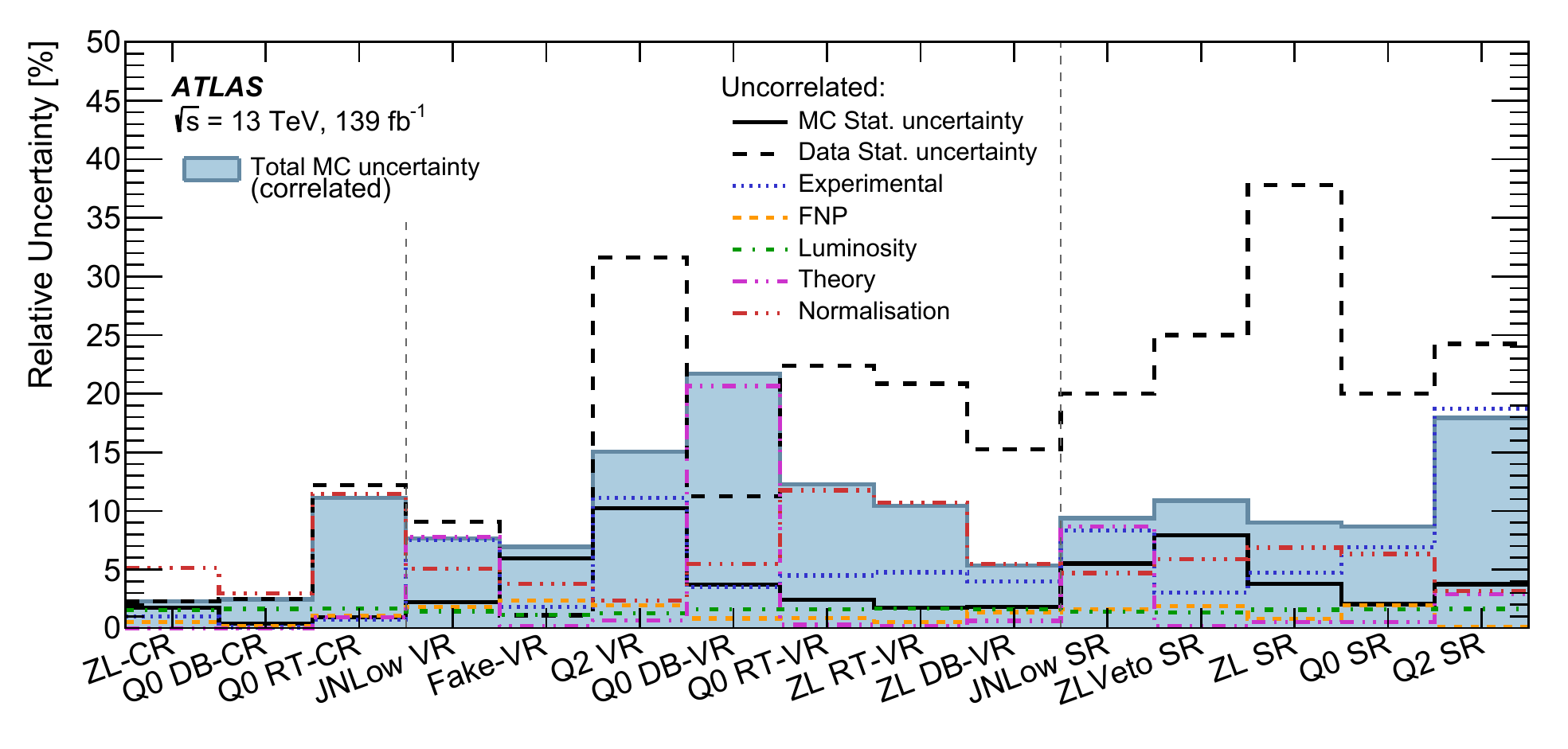}
\caption{Relative contributions from different sources of statistical
and systematic uncertainty to the total background yield estimates
after the fit. \textit{Experimental} uncertainties are related to
the lepton, jet and \MET selection and reconstruction, and also to lepton
charge misidentification.
\textit{FNP} includes the fake or non-prompt leptons contribution. \textit{Luminosity} is related to the
luminosity uncertainty that affects the background simulation
yields. \textit{Theory} includes theoretical uncertainties
associated with the PDF, \alphas, and renormalisation
and factorisation scales. \textit{Normalisation} is related to the
diboson and rare top-quark normalisation factors extracted by the
likelihood fit. Systematic uncertainties are calculated by changing each nuisance parameter from its fit value by one standard deviation,
keeping all the other parameters at their central values, and comparing
the resulting event yield with the nominal yield.
Individual uncertainties can be correlated within each region, and do not necessarily add
in quadrature to the total background uncertainty, which is shown as
\textit{Total MC uncertainty (correlated)}. \textit{Data Stat.\ uncertainty} refers to the statistical uncertainty of the collected data. }
\label{fig:34RegionsSys}
\end{figure}

The relative uncertainties in the background yield estimates are
shown in  \Fig{\ref{fig:34RegionsSys}} for all analysis
regions in the three- and four-lepton channels. The dominant uncertainty in the SRs, and in most of the other regions, is
the statistical uncertainty of the data, which varies
from $\SI{20}{\%}$ to $\SI{37}{\%}$ depending on the signal region. The MC statistical uncertainty varies from $\SI{2}{\%}$ to $\SI{7}{\%}$ instead.
In the Q2~SR an uncertainty contribution close to the data statistical uncertainty comes from the charge misidentification background, considered in the \textit{Experimental} category.

In the absence of a significant deviation from SM expectations,
\SI{95}{\%} CL upper limits on the  signal
production cross-section are derived using the \(
\mathrm{CL}_\mathrm{s} \)
method~\cite{Read:2002hq}. The  upper limits on the production cross-sections of the \( pp \rightarrow W^{*} \rightarrow \Nz \Lpm \)
and
\( pp \rightarrow Z^{*} \rightarrow \Lpm \Lmp \) processes
are evaluated as a function of the heavy-lepton
mass, using the three- and four-lepton channels with the democratic $\mathcal{B}_\ell$ scenario.
By comparing the upper limits on the cross-section with the
theoretical cross-section calculation as a function of the
heavy-lepton mass, a lower limit on the
mass of the \typeIIIseesaw heavy leptons \Nz and \Lpm is derived.
The observed  (expected) exclusion
limit is \SI{870}{\GeV} ( $900^{+80}_{-80}$\,GeV).

\begin{figure}[tbp]
\centering
\includegraphics[width=0.9\textwidth]{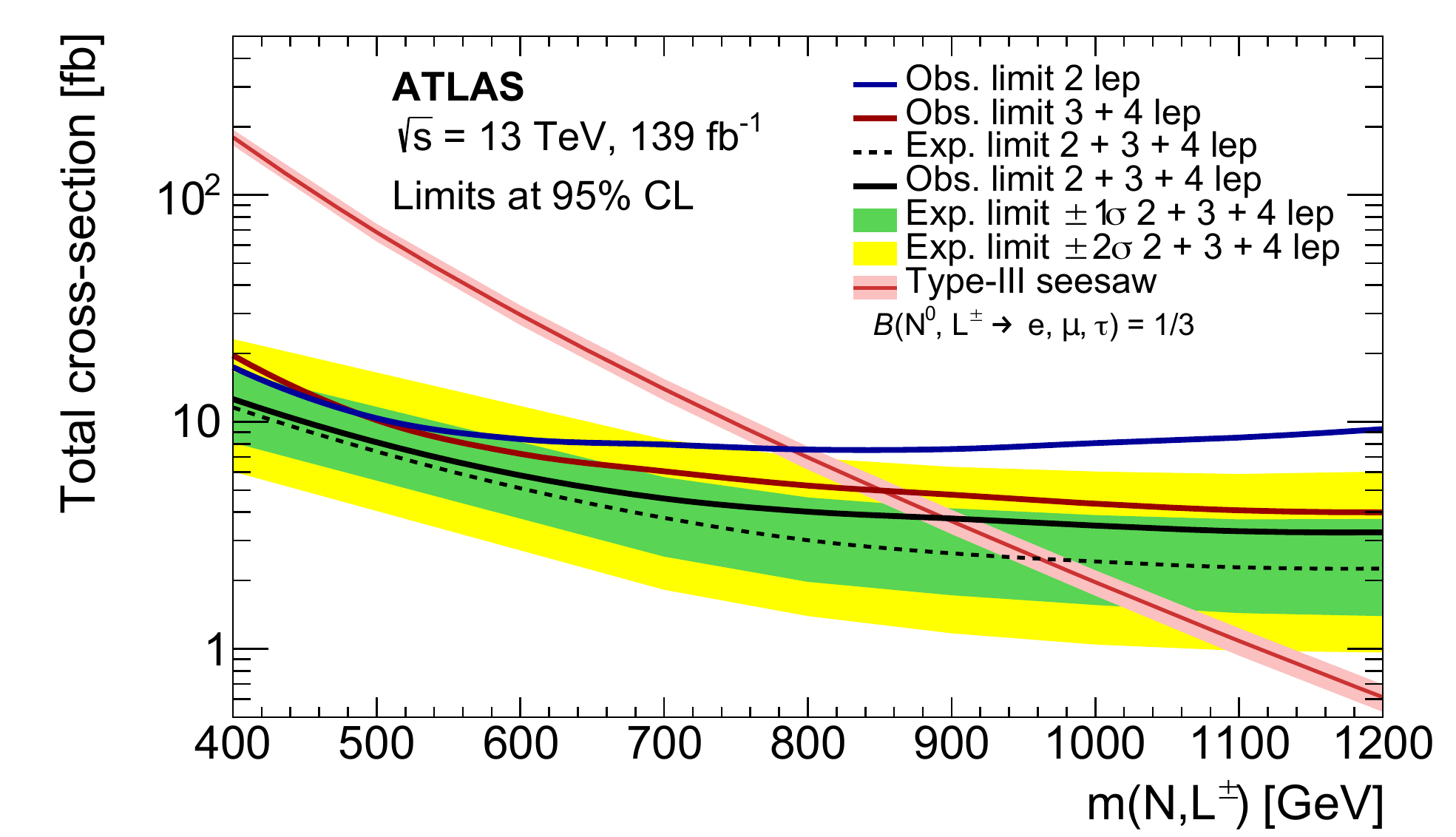}
\caption{Expected and observed
exclusion limits in the two-lepton channel
(from Ref.~\cite{EXOT-2018-33}), the  three- and four-lepton channels, and
the two-, three- and four-lepton channels for the
\typeIIIseesaw process with the corresponding
one- and two-standard-deviation uncertainty bands, showing the \SI{95}{\%} CL upper limit
on the cross-section.
The theoretical signal cross-section prediction,
given by the NLO calculation~\cite{Fuks:2012qx,Fuks:2013vua},
with its corresponding uncertainty band is also shown.
}
\label{fig:exclusion_234L}
\end{figure}

\begin{figure}[tbp]
\centering
\includegraphics[width=0.9\textwidth]{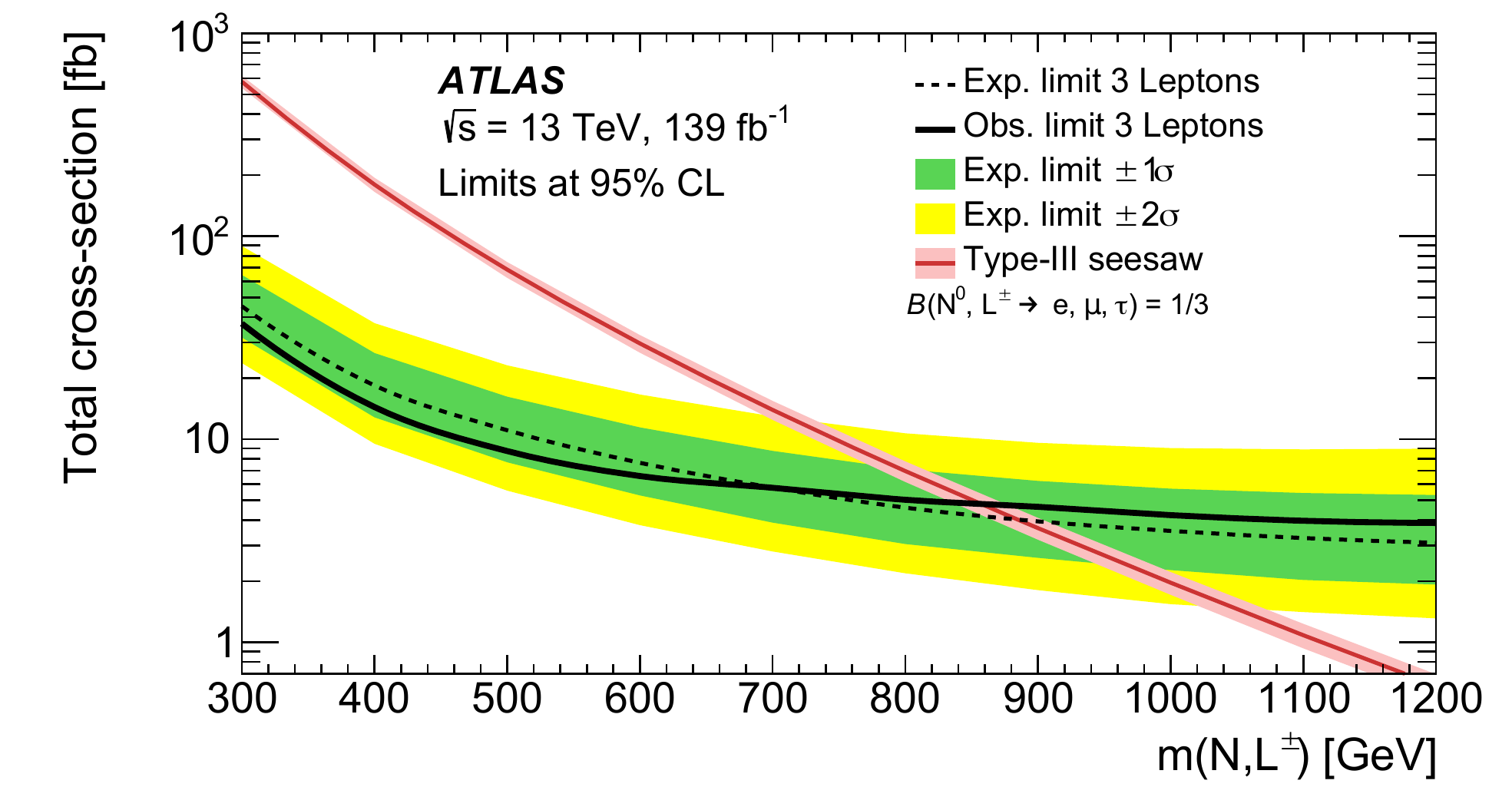}
\caption{Expected and observed \SI{95}{\%} \( \mathrm{CL}_\mathrm{s} \)
exclusion limits in the three lepton channel for the \typeIIIseesaw process with the corresponding
one- and two-standard-deviation bands, showing the \SI{95}{\%} CL upper limit
on the cross-section.
The theoretical signal cross-section prediction,
given by the NLO calculation~\cite{Fuks:2012qx,Fuks:2013vua}, is shown
with the corresponding uncertainty bands for the expected limit.
}
\label{fig:exclusion_3L}
\end{figure}

\begin{figure}[tbp]
\centering
\includegraphics[width=0.9\textwidth]{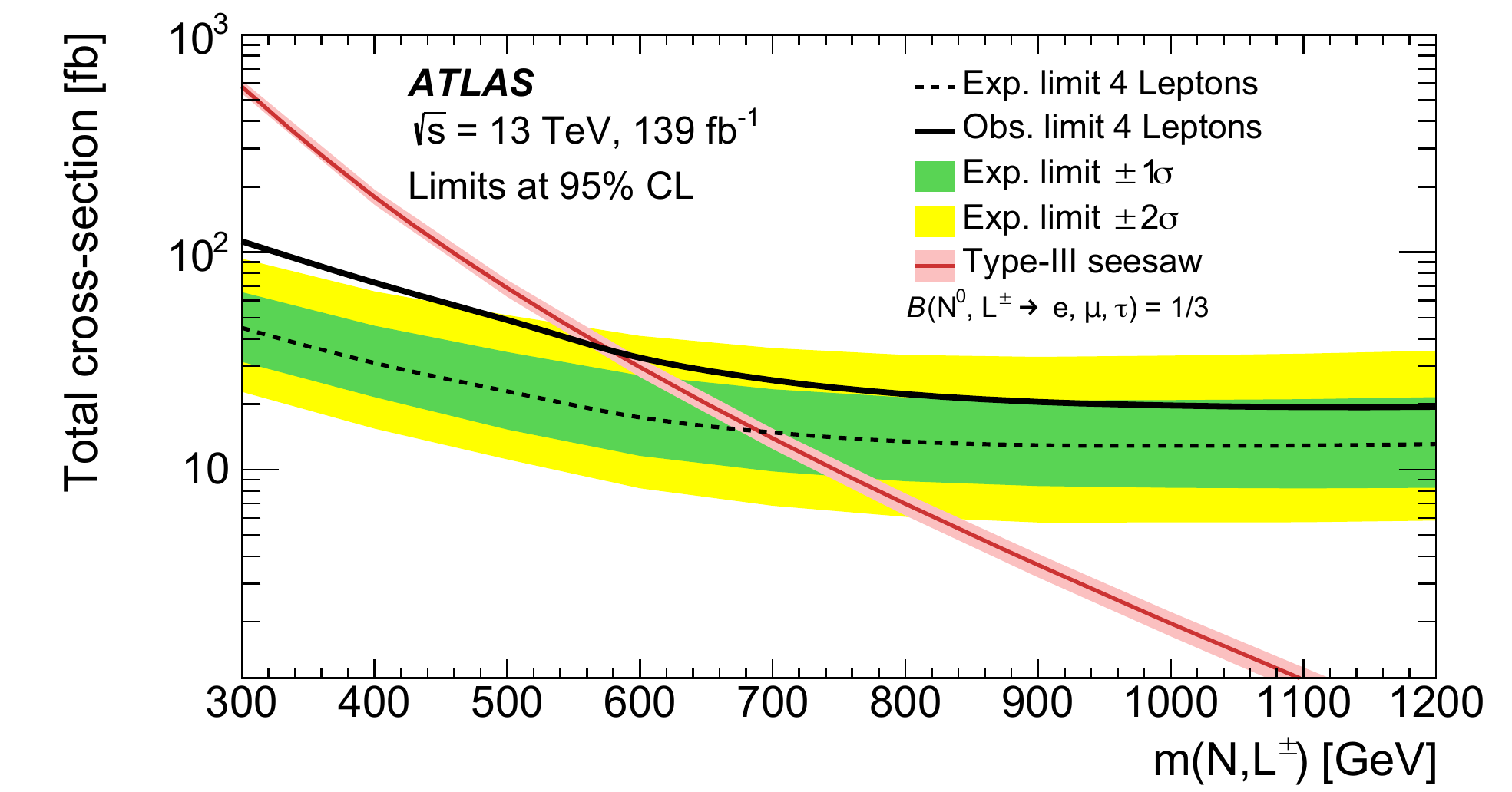}
\caption{Expected and observed \SI{95}{\%} \( \mathrm{CL}_\mathrm{s} \)
exclusion limits in the four lepton channel for the \typeIIIseesaw process with the corresponding
one- and two-standard-deviation bands, showing the \SI{95}{\%} CL upper limit
on the cross-section.
The theoretical signal cross-section prediction,
given by the NLO calculation~\cite{Fuks:2012qx,Fuks:2013vua}, is shown
with the corresponding uncertainty bands for the expected limit.
}
\label{fig:exclusion_4L}
\end{figure}

The signal hypothesis in the three- and four-lepton channel result is also
tested in a combined fit with the
similar \typeIIIseesaw search regions in the two-lepton
channel~\cite{EXOT-2018-33}. All the CRs, VRs and SRs in the various lepton
multiplicity regions are statistically independent. The reconstruction algorithms and working points are the same in all cases, and the
FNP and lepton charge misidentification backgrounds are estimated using the same method.
The parameter of interest, namely the
number of signal events,
and common systematic
uncertainties are treated as correlated. Normalisations of the diboson, \ttbar (for the
two-lepton multiplicity region) and rare top-quark (for the three- and four-lepton multiplicity regions) backgrounds are  treated as
uncorrelated since they
account for different physics processes and different acceptances in
each final state. The three-lepton channel's limit dominates in the high
heavy-lepton mass region, while the two-lepton channel dominates in the
lower mass region.
The combined observed (expected) exclusion limits on the total cross-section are shown in \Fig{\ref{fig:exclusion_234L}}, excluding heavy-lepton masses lower
than \SI{910}{\GeV} ( $960^{+90}_{-80}$\,GeV) at 95\% CL. The combined observed (expected) exclusion limits on the total cross-section restraining to three-lepton and four-lepton channels are shown in \Fig{\ref{fig:exclusion_3L} and \ref{fig:exclusion_4L}} respectively.

\FloatBarrier

\section{Conclusion}
\label{sec:conclusion}
ATLAS has searched for pair-produced heavy leptons predicted by the
\typeIIIseesaw model in \SI{139}{\ifb} of data from proton--proton collisions at
\( \sqrt{s} = \SI{13}{\TeV} \), recorded during the 2015--2018
data-taking period.
A lower limit on the
mass of the \typeIIIseesaw heavy leptons \Nz and \Lpm is derived for final states with three or four light leptons. No significant deviation from SM expectations is observed. The observed  (expected) exclusion
limit on the heavy-lepton mass is \SI{870}{\GeV} ($900^{+80}_{-80}$\,GeV) at the 95\% CL.
This result is combined with the result of
the two-lepton analysis, which used very similar experimental
methodologies and treatment of statistics.
In the full combination, heavy leptons with masses below
\SI{910}{\GeV} are excluded at the
\SI{95}{\%} CL, while the expected lower limit on the mass is
\SI[parse-numbers=false]{960^{+90}_{-80}}{\GeV}. This is the
most stringent limit to date on the  \typeIIIseesaw model from events with light leptons at LHC\@.

\clearpage
 
\section*{Acknowledgements}


We thank CERN for the very successful operation of the LHC, as well as the
support staff from our institutions without whom ATLAS could not be
operated efficiently.
 
We acknowledge the support of
ANPCyT, Argentina;
YerPhI, Armenia;
ARC, Australia;
BMWFW and FWF, Austria;
ANAS, Azerbaijan;
SSTC, Belarus;
CNPq and FAPESP, Brazil;
NSERC, NRC and CFI, Canada;
CERN;
ANID, Chile;
CAS, MOST and NSFC, China;
Minciencias, Colombia;
MEYS CR, Czech Republic;
DNRF and DNSRC, Denmark;
IN2P3-CNRS and CEA-DRF/IRFU, France;
SRNSFG, Georgia;
BMBF, HGF and MPG, Germany;
GSRI, Greece;
RGC and Hong Kong SAR, China;
ISF and Benoziyo Center, Israel;
INFN, Italy;
MEXT and JSPS, Japan;
CNRST, Morocco;
NWO, Netherlands;
RCN, Norway;
MEiN, Poland;
FCT, Portugal;
MNE/IFA, Romania;
JINR;
MES of Russia and NRC KI, Russian Federation;
MESTD, Serbia;
MSSR, Slovakia;
ARRS and MIZ\v{S}, Slovenia;
DSI/NRF, South Africa;
MICINN, Spain;
SRC and Wallenberg Foundation, Sweden;
SERI, SNSF and Cantons of Bern and Geneva, Switzerland;
MOST, Taiwan;
TENMAK, T\"urkiye;
STFC, United Kingdom;
DOE and NSF, United States of America.
In addition, individual groups and members have received support from
BCKDF, CANARIE, Compute Canada and CRC, Canada;
COST, ERC, ERDF, Horizon 2020 and Marie Sk{\l}odowska-Curie Actions, European Union;
Investissements d'Avenir Labex, Investissements d'Avenir Idex and ANR, France;
DFG and AvH Foundation, Germany;
Herakleitos, Thales and Aristeia programmes co-financed by EU-ESF and the Greek NSRF, Greece;
BSF-NSF and GIF, Israel;
Norwegian Financial Mechanism 2014-2021, Norway;
NCN and NAWA, Poland;
La Caixa Banking Foundation, CERCA Programme Generalitat de Catalunya and PROMETEO and GenT Programmes Generalitat Valenciana, Spain;
G\"{o}ran Gustafssons Stiftelse, Sweden;
The Royal Society and Leverhulme Trust, United Kingdom.
 
The crucial computing support from all WLCG partners is acknowledged gratefully, in particular from CERN, the ATLAS Tier-1 facilities at TRIUMF (Canada), NDGF (Denmark, Norway, Sweden), CC-IN2P3 (France), KIT/GridKA (Germany), INFN-CNAF (Italy), NL-T1 (Netherlands), PIC (Spain), ASGC (Taiwan), RAL (UK) and BNL (USA), the Tier-2 facilities worldwide and large non-WLCG resource providers. Major contributors of computing resources are listed in Ref.~\cite{ATL-SOFT-PUB-2021-003}.
 

\clearpage
\printbibliography
 
\clearpage
 
\begin{flushleft}
\hypersetup{urlcolor=black}
{\Large The ATLAS Collaboration}

\bigskip

\AtlasOrcid[0000-0002-6665-4934]{G.~Aad}$^\textrm{\scriptsize 99}$,    
\AtlasOrcid[0000-0002-5888-2734]{B.~Abbott}$^\textrm{\scriptsize 124}$,    
\AtlasOrcid[0000-0002-7248-3203]{D.C.~Abbott}$^\textrm{\scriptsize 100}$,    
\AtlasOrcid[0000-0002-2788-3822]{A.~Abed~Abud}$^\textrm{\scriptsize 34}$,    
\AtlasOrcid[0000-0002-1002-1652]{K.~Abeling}$^\textrm{\scriptsize 51}$,    
\AtlasOrcid[0000-0002-2987-4006]{D.K.~Abhayasinghe}$^\textrm{\scriptsize 91}$,    
\AtlasOrcid[0000-0002-8496-9294]{S.H.~Abidi}$^\textrm{\scriptsize 27}$,    
\AtlasOrcid[0000-0001-5329-6640]{H.~Abramowicz}$^\textrm{\scriptsize 157}$,    
\AtlasOrcid[0000-0002-1599-2896]{H.~Abreu}$^\textrm{\scriptsize 156}$,    
\AtlasOrcid[0000-0003-0403-3697]{Y.~Abulaiti}$^\textrm{\scriptsize 5}$
% ...
\bigskip
\\

% ...
$^{5}$High Energy Physics Division, Argonne National Laboratory, Argonne IL; United States of America.\\
% ...
$^{27}$Physics Department, Brookhaven National Laboratory, Upton NY; United States of America.\\
% ...
$^{34}$CERN, Geneva; Switzerland.\\
% ...
$^{51}$II. Physikalisches Institut, Georg-August-Universit\"{a}t G\"ottingen, G\"ottingen; Germany.\\
% ...
$^{91}$Department of Physics, Royal Holloway University of London, Egham; United Kingdom.\\
% ...
$^{99}$CPPM, Aix-Marseille Universit\'e, CNRS/IN2P3, Marseille; France.\\
$^{100}$Department of Physics, University of Massachusetts, Amherst MA; United States of America.\\
% ...
$^{124}$Homer L. Dodge Department of Physics and Astronomy, University of Oklahoma, Norman OK; United States of America.\\
% ...
$^{156}$Department of Physics, Technion, Israel Institute of Technology, Haifa; Israel.\\
$^{157}$Raymond and Beverly Sackler School of Physics and Astronomy, Tel Aviv University, Tel Aviv; Israel.\\
% ...

\end{flushleft}


\end{document}